\documentclass[11pt,a4paper]{article}
\usepackage{jcappub}
\usepackage{booktabs} 
\usepackage{bbold}

\newcommand{\bfeta}{\boldsymbol\eta}
\newcommand{\beqra}{\begin{eqnarray}}
\newcommand{\eeqra}{\end{eqnarray}}
\newcommand{\beq}{\begin{equation}}
\newcommand{\eeq}{\end{equation}}

\title{Global limits and interference patterns in dark matter direct detection}
\author[a]{Riccardo Catena}
\author[b]{and Paolo Gondolo}
\affiliation[a]{Institut f\"ur Theoretische Physik, Friedrich-Hund-Platz 1, 37077 G\"ottingen, Germany}
\affiliation[b]{Department of Physics and Astronomy, University of Utah, 115 South 1400 East \#201, Salt Lake City, UT 84112, USA}
\emailAdd{riccardo.catena@theorie.physik.uni-goettingen.de}
\emailAdd{paolo.gondolo@utah.edu}

\abstract{We compare the general effective theory of one-body dark matter nucleon interactions to current direct detection experiments in a global multidimensional statistical analysis. We derive exclusion limits on the 28 isoscalar and isovector coupling constants of the theory, and show that current data place interesting constraints on dark matter-nucleon interaction operators usually neglected in this context. We characterize the interference patterns that can arise in dark matter direct detection from pairs of dark matter-nucleon interaction operators, or from isoscalar and isovector components of the same operator.
We find that commonly neglected destructive interference effects weaken standard direct detection exclusion limits by up to one order of magnitude in the coupling constants.}


\begin{document}
\maketitle

\section{Introduction}
The simultaneous operation of dark matter direct detection experiments exploiting complementary techniques, and target materials, raise the issue of how to effectively interpret the information gathered on dark matter~\cite{Ahmed:2009zw,Ahmed:2010wy,Agnese:2014aze,Agnese:2013jaa,Aprile:2012nq,Angle:2011th,Akerib:2013tjd,Behnke:2012ys,Archambault:2012pm}. 

The currently favored strategy in the analysis of dark matter direct detection experiments relies on drastic simplifying assumptions regarding the underlying dark matter-nucleon interaction~\cite{Lewin:1995rx,Catena:2013pka}. For instance, the dark matter-nucleon interaction is often assumed to be independent of the momentum transferred in the scattering, and of the dark matter-nucleus relative velocity. Moreover, the possible interference of different dark matter-nucleon interactions is commonly neglected in the data analysis. At the same time, distinct experiments are usually interpreted separately, even when they are compatible and, most importantly, complementary.

Though this approach is a well motivated first approximation, the substantial progress recently made in the field, and the efforts planned for the next decade, motivate the exploration of more sophisticated strategies. 

In this work we explore two extensions of the standard approach to dark matter direct detection. Firstly, we interpret current dark matter direct detection experiments without assuming the knowledge of the dark matter-nucleon interaction a priori. In other words, we do not artificially restrict the analysis to constant spin-independent or spin-dependent interactions. In contrast, we interpret current observations within the general effective theory of one-body dark matter nucleon interactions, which predicts 28 independent isoscalar and isovector dark matter-nucleon interaction types. Secondly, we interpret current direct detection experiments globally~\cite{Kopp:2009qt,Catena:2014uqa}, i.e. varying all correlated coupling constants simultaneously, and considering several direct detection experiments in a single multidimensional Likelihood analysis. 

Our approach allows to characterize the interference patterns arising in dark matter direct detection from pairs of dark matter-nucleon interaction operators, or from isoscalar and isovector components of a given operator. 
We refer to these interference patterns as multi-interaction interference effects. A key result of this work is to quantify the impact of multi-interaction interference effects on the calculation of dark matter direct detection exclusion limits. 

Effective theories for dark matter-nucleon interactions were already explored in the context of dark matter direct detection in~\cite{Chang:2009yt,Fan:2010gt,Fornengo:2011sz,Fitzpatrick:2012ix,Fitzpatrick:2012ib,Menendez:2012tm,Cirigliano:2012pq,Anand:2013yka,DelNobile:2013sia,Klos:2013rwa,Peter:2013aha,Hill:2013hoa,Catena:2014uqa,Catena:2014hla,Catena:2014epa,Gluscevic:2014vga,Panci:2014gga,Vietze:2014vsa,Barello:2014uda,Schneck:2015eqa,Hoferichter:2015ipa}, in the analysis of neutrino telescope observations in~\cite{Guo:2013ypa,Liang:2013dsa,Blumenthal:2014cwa,Catena:2015uha,Catena:2015iea}, and in the context of helioseismology in~\cite{Vincent:2013lua,Lopes:2014aoa,Vincent:2014jia,Vincent:2015gqa}. The prospects for dark matter direct detection in general effective theories were first investigated in~\cite{Catena:2014epa,Catena:2014hla,Gluscevic:2014vga}, and later in~\cite{Schneck:2015eqa}. Ref.~\cite{Catena:2014hla} also quantifies the bias in the dark matter particle mass and cross-section reconstruction induced by fitting observations assuming constant dark matter-nucleon interactions, when dark matter interacts with nucleons differently. 
The importance of multi-interaction interference effects in dark matter direct detection was first noticed in~\cite{DelNobile:2011yb,Catena:2014uqa,Hamze:2014wca}, and recently also in~\cite{Schneck:2015eqa}.

The paper is organized as follows. In Sec.~\ref{sec:EFT} we review the non-relativistic effective theory of dark matter scattering from nucleons and nuclei. Sec.~\ref{sec:statistics} is devoted to the statistical framework adopted in the analysis of the dark matter direct detection experiments introduced in Sec.~\ref{sec:exp}. We present our results in Sec.~\ref{sec:results}, and conclude in Sec.~\ref{sec:conclusions}. Important equations are listed in Appendix~\ref{sec:appDM}.

\section{Dark matter-nucleus scattering in effective theories}
\label{sec:EFT}
In this section we review the non-relativistic effective theory of dark matter scattering from nucleons and nuclei. 

\subsection{Interaction Hamiltonian}
Assuming one-body dark matter-nucleon interactions mediated by a heavy spin-1 or spin-0 particle, the most general Hamiltonian density for dark matter-nucleus scattering is given by~\cite{Fitzpatrick:2012ix}
\begin{equation}
\hat{\mathcal{H}}_{\rm T}= \sum_{i=1}^{A}  \sum_{\tau=0,1} \sum_{k} c_k^{\tau}\hat{\mathcal{O}}_{k}^{(i)} \, t^{\tau}_{(i)} \,.
\label{eq:H_I}
\end{equation}
It is the sum of $A$ terms, one for each nucleon in the target nucleus. The $A$ terms in the sum are linear combinations of 14 Galilean invariant quantum mechanical interaction operators $\hat{\mathcal{O}}_k^{(i)}$, $k=1,3,\dots,15$. They are listed in Tab.~\ref{tab:operators}. 

The 14 operators $\hat{\mathcal{O}}_k^{(i)}$ are constructed from the momentum transfer operator ${\bf{\hat{q}}}$, the relative transverse velocity operator ${\bf{\hat{v}}}^{\perp}$, and the dark matter particle and nucleon spin operators, ${\bf{\hat{S}}}_\chi$ and ${\bf{\hat{S}}}_N$, respectively~\cite{Fitzpatrick:2012ix}. 
In Eq.~(\ref{eq:H_I}) $t^{0}= \mathbb{1}$ is the identity in isospin space, $t^{1}= \tau_3$ is the third Pauli matrix, and $m_N$ is the nucleon mass. The isoscalar and isovector coupling constants, respectively $c_k^{0}$ and $c_k^{1}$, are related to the coupling constants for protons and neutrons by $c^{p}_k=(c^{0}_k+c^{1}_k)/2$, and $c^{n}_k=(c^{0}_k-c^{1}_k)/2$. They have dimension mass to the power of $-2$. 

The Hamiltonian density in Eq.~(\ref{eq:H_I}) admits the following coordinate space representation~\cite{Fitzpatrick:2012ix} 
\begin{eqnarray}
\hat{\mathcal{H}}_{\rm T}({\bf{r}}) = \sum_{\tau=0,1} &\Bigg\{&
\sum_{i=1}^A  \hat{l}_0^{\tau}~ \delta({\bf{r}}-{\bf{r}}_i)
 + \sum_{i=1}^A  {\bf{\hat{l}}}_5^{\tau} \cdot \vec{\sigma}_i \,\delta({\bf{r}}-{\bf{r}}_i) \nonumber\\
 &+&   \sum_{i=1}^A {\bf{\hat{l}}}_M^{\tau} \cdot \frac{1}{2 m_N} \Bigg[i \overleftarrow{\nabla}_{\bf{r}}\delta({\bf{r}}-{\bf{r}}_i) -i \delta({\bf{r}}-{\bf{r}}_i)\overrightarrow{\nabla}_{\bf{r}} \Bigg]  \nonumber \\
&+& \sum_{i=1}^A {\bf{\hat{l}}}_E^{\tau} \cdot \frac{1}{2m_N} \Bigg[ \overleftarrow{\nabla}_{\bf{r}} \times \vec{\sigma}_i \,\delta({\bf{r}}-{\bf{r}}_i) +\delta({\bf{r}}-{\bf{r}}_i)\
  \vec{\sigma}_i \times \overrightarrow{\nabla}_{\bf{r}} \Bigg] \Bigg\} t^{\tau}_{(i)}\nonumber\\
\label{eq:Hx}
\end{eqnarray}
where $\vec{\sigma}_i$ denotes the set of three Pauli matrices representing the spin operator of the $i$th-nucleon in the target nucleus. ${\bf{r}}_i$ is the $i$th-nucleon position in the nucleus center of mass frame, and 
\begin{eqnarray}
\label{eq:ls}
\hat{l}_0^\tau &=& c_1^\tau + i  \left( {{\bf{\hat{q}}} \over m_N}  \times {\bf{\hat{v}}}_{T}^\perp \right) \cdot  {\bf{\hat{S}}}_\chi  ~c_5^\tau
+ {\bf{\hat{v}}}_{T}^\perp \cdot {\bf{\hat{S}}}_\chi  ~c_8^\tau + i {{\bf{\hat{q}}} \over m_N} \cdot {\bf{\hat{S}}}_\chi ~c_{11}^\tau \nonumber \\
{\bf{\hat{l}}}_5^{\tau}&=& {1 \over 2} \left[ i {{\bf{\hat{q}}} \over m_N} \times {\bf{\hat{v}}}_{T}^\perp~ c_3^\tau + {\bf{\hat{S}}}_\chi ~c_4^\tau
+  {{\bf{\hat{q}}} \over m_N}~{{\bf{\hat{q}}} \over m_N} \cdot {\bf{\hat{S}}}_\chi ~c_6^\tau
+   {\bf{\hat{v}}}_{T}^\perp ~c_7^\tau + i {{\bf{\hat{q}}} \over m_N} \times {\bf{\hat{S}}}_\chi ~c_9^\tau + i {{\bf{\hat{q}}} \over m_N}~c_{10}^\tau \right. \nonumber \\
&+&  \left. {\bf{\hat{v}}}_{T}^\perp \times {\bf{\hat{S}}}_\chi ~c_{12}^\tau
+i  {{\bf{\hat{q}}} \over m_N} {\bf{\hat{v}}}_{T}^\perp \cdot {\bf{\hat{S}}}_\chi ~c_{13}^\tau+i {\bf{\hat{v}}}_{T}^\perp {{\bf{\hat{q}}} \over m_N} \cdot {\bf{\hat{S}}}_\chi ~ c_{14}^\tau+{{\bf{\hat{q}}} \over\
 m_N} \times {\bf{\hat{v}}}_{T}^\perp~ {{\bf{\hat{q}}} \over m_N} \cdot {\bf{\hat{S}}}_\chi ~ c_{15}^\tau  \right]\nonumber \\
{\bf{\hat{l}}}_M^{\tau} &=&   i {{\bf{\hat{q}}} \over m_N}  \times {\bf{\hat{S}}}_\chi ~c_5^\tau - {\bf{\hat{S}}}_\chi ~c_8^\tau \nonumber \\
{\bf{\hat{l}}}_E^{\tau} &=& {1 \over 2} \left[  {{\bf{\hat{q}}} \over m_N} ~ c_3^\tau +i {\bf{\hat{S}}}_\chi~c_{12}^\tau - {{\bf{\hat{q}}} \over  m_N} \times{\bf{\hat{S}}}_\chi  ~c_{13}^\tau-i 
{{\bf{\hat{q}}} \over  m_N} {{\bf{\hat{q}}} \over m_N} \cdot {\bf{\hat{S}}}_\chi  ~c_{15}^\tau \right] \,. \nonumber\\
\end{eqnarray}
In the last expression ${\bf{\hat{v}}}^{\perp}_T={\bf{\hat{v}}}^{\perp} - {\bf{\hat{v}}}^{\perp}_N$, where ${\bf{\hat{v}}}^{\perp}_N$ is an operator acting on the $i$th-nucleon space coordinate. Explicit coordinate space representations for the operators ${\bf{\hat{q}}}$, ${\bf{\hat{v}}}^{\perp}_T$, and ${\bf{\hat{v}}}^{\perp}_N$ can be found in \cite{Catena:2015uha}.

\begin{table}[t]
    \centering
    \begin{tabular}{ll}
    \toprule
        $\hat{\mathcal{O}}_1 = \mathbb{1}_{\chi N}$ & $\hat{\mathcal{O}}_9 = i{\bf{\hat{S}}}_\chi\cdot\left(\hat{{\bf{S}}}_N\times\frac{{\bf{\hat{q}}}}{m_N}\right)$  \\
        $\hat{\mathcal{O}}_3 = i\hat{{\bf{S}}}_N\cdot\left(\frac{{\bf{\hat{q}}}}{m_N}\times{\bf{\hat{v}}}^{\perp}\right)$ \hspace{2 cm} &   $\hat{\mathcal{O}}_{10} = i\hat{{\bf{S}}}_N\cdot\frac{{\bf{\hat{q}}}}{m_N}$   \\
        $\hat{\mathcal{O}}_4 = \hat{{\bf{S}}}_{\chi}\cdot \hat{{\bf{S}}}_{N}$ &   $\hat{\mathcal{O}}_{11} = i{\bf{\hat{S}}}_\chi\cdot\frac{{\bf{\hat{q}}}}{m_N}$   \\                                                                             
        $\hat{\mathcal{O}}_5 = i{\bf{\hat{S}}}_\chi\cdot\left(\frac{{\bf{\hat{q}}}}{m_N}\times{\bf{\hat{v}}}^{\perp}\right)$ &  $\hat{\mathcal{O}}_{12} = \hat{{\bf{S}}}_{\chi}\cdot \left(\hat{{\bf{S}}}_{N} \times{\bf{\hat{v}}}^{\perp} \right)$ \\                                                                                                                 
        $\hat{\mathcal{O}}_6 = \left({\bf{\hat{S}}}_\chi\cdot\frac{{\bf{\hat{q}}}}{m_N}\right) \left(\hat{{\bf{S}}}_N\cdot\frac{\hat{{\bf{q}}}}{m_N}\right)$ &  $\hat{\mathcal{O}}_{13} =i \left(\hat{{\bf{S}}}_{\chi}\cdot {\bf{\hat{v}}}^{\perp}\right)\left(\hat{{\bf{S}}}_{N}\cdot \frac{{\bf{\hat{q}}}}{m_N}\right)$ \\   
        $\hat{\mathcal{O}}_7 = \hat{{\bf{S}}}_{N}\cdot {\bf{\hat{v}}}^{\perp}$ &  $\hat{\mathcal{O}}_{14} = i\left(\hat{{\bf{S}}}_{\chi}\cdot \frac{{\bf{\hat{q}}}}{m_N}\right)\left(\hat{{\bf{S}}}_{N}\cdot {\bf{\hat{v}}}^{\perp}\right)$  \\
        $\hat{\mathcal{O}}_8 = \hat{{\bf{S}}}_{\chi}\cdot {\bf{\hat{v}}}^{\perp}$  & $\hat{\mathcal{O}}_{15} = -\left(\hat{{\bf{S}}}_{\chi}\cdot \frac{{\bf{\hat{q}}}}{m_N}\right)\left[ \left(\hat{{\bf{S}}}_{N}\times {\bf{\hat{v}}}^{\perp} \right) \cdot \frac{{\bf{\hat{q}}}}{m_N}\right] $ \\                                                                               
    \bottomrule
    \end{tabular}
    \caption{Non-relativistic quantum mechanical operators defining the general effective theory of one-body dark matter-nucleon interactions. Because of the nucleon mass, $m_N$, in the equations above all operators have the same mass dimension. For simplicity, here and in the next sections we omit the nucleon index $(i)$ adopted in Sec.~\ref{sec:EFT}.} 
    \label{tab:operators}
\end{table}

Eq.~(\ref{eq:Hx}) shows that dark matter couples to the constituent nucleons through  the nuclear vector charge and spin current (first line), the nuclear convection current (second line), and the nuclear spin-velocity current (last line).  In Eq.~(\ref{eq:Hx}) we have omitted the nuclear axial charge, since it does not contribute to dark matter-nucleus scattering cross-sections when nuclear ground states are eigenstates of $P$ and $CP$, as it is commonly assumed for target nuclei in dark matter direct detection.

\subsection{Transition amplitude}
We use Eq.~(\ref{eq:Hx}) to calculate $\langle |\mathcal{M}_{NR}|^2\rangle_{\rm spins}$, i.e. the square modulus of the amplitude for dark matter-nucleus scattering. Assuming that nuclear ground states are eigenstates of $P$ and $CP$, and averaging (summing) over initial (final) spin configurations, one finds\footnote{Strictly speaking $M_{NR}$ is not an amplitude, as with our definition of $c_k^\tau$ it has dimension mass$^{-4}$.} 
\begin{align}
\langle |\mathcal{M}_{NR}|^2\rangle_{\rm spins} =  \frac{4\pi}{2J+1}\sum_{\tau,\tau'} &\bigg[ \sum_{k=M,\Sigma',\Sigma''} R^{\tau\tau'}_k\left(v_T^{\perp 2}, {q^2 \over m_N^2} \right) W_k^{\tau\tau'}(y) \nonumber\\
&+{q^{2} \over m_N^2} \sum_{k=\Phi'', \Phi'' M, \tilde{\Phi}', \Delta, \Delta \Sigma'} R^{\tau\tau'}_k\left(v_T^{\perp 2}, {q^2 \over m_N^2}\right) W_k^{\tau\tau'}(y) \bigg] \,. \nonumber\\
\label{eq:M}
\end{align}
The index $k$ extends over the nuclear response functions $W_k^{\tau\tau'}(y)$ defined below in Eq.~(\ref{eq:W}).
Assuming the harmonic oscillator basis for single-nucleon states, $y=(bq/2)^2$ with
\begin{align}
b=\sqrt{41.467/(45 A^{-1/3}-25A^{-2/3})}~{\rm fm}\,.
\end{align} 
The 8 dark matter response functions $R^{\tau\tau'}_k$ in Eq.~(\ref{eq:M}) depend on matrix elements of the operators in (\ref{eq:ls}), and are listed in Appendix~\ref{sec:appDM}. They are functions of the dark matter particle spin, of $q^2/m_N^2$, and of 
\begin{equation}
v_T^{\perp\,2}=v^2-q^2/(4\mu_T^2)\,, 
\end{equation}
where $\mu_T$ and $v$ are the reduced dark matter-nucleus mass and the dark matter-nucleus relative velocity, respectively.
The 8 nuclear response functions $W_k^{\tau\tau'}(y)$ in Eq.~(\ref{eq:M}) are quadratic in nuclear matrix elements, and defined as follows
\begin{equation}
W_{AB}^{\tau \tau^\prime}(y)= \sum_{L}  \langle J,T,M_T ||~ A_{L;\tau} (q)~ || J,T,M_T \rangle \langle J,T,M_T ||~ B_{L;\tau^\prime} (q)~ || J,T,M_T \rangle \,,
\label{eq:W}
\end{equation}
where $|J,T,M_T \rangle$ represents a nuclear state of spin $J$, isospin $T$, and isospin magnetic quantum number $M_T$. The reduction operation, i.e. $||\cdot||$, is done via the Wigner-Eckart theorem. 
In Eq.~(\ref{eq:W}), $A$ and $B$ can each be one of the following nuclear response operators
\begin{eqnarray}
M_{LM;\tau}(q) &=& \sum_{i=1}^{A} M_{LM}(q {\bf{r}}_i) t^{\tau}_{(i)}\nonumber\\
\Sigma'_{LM;\tau}(q) &=& -i \sum_{i=1}^{A} \left[ \frac{1}{q} \overrightarrow{\nabla}_{{\bf{r}}_i} \times {\bf{M}}_{LL}^{M}(q {\bf{r}}_i)  \right] \cdot \vec{\sigma}_i \, t^{\tau}_{(i)}\nonumber\\
\Sigma''_{LM;\tau}(q) &=&\sum_{i=1}^{A} \left[ \frac{1}{q} \overrightarrow{\nabla}_{{\bf{r}}_i} M_{LM}(q {\bf{r}}_i)  \right] \cdot \vec{\sigma}_i \, t^{\tau}_{(i)}\nonumber\\
\Delta_{LM;\tau}(q) &=&\sum_{i=1}^{A}  {\bf{M}}_{LL}^{M}(q {\bf{r}}_i) \cdot \frac{1}{q}\overrightarrow{\nabla}_{{\bf{r}}_i} t^{\tau}_{(i)} \nonumber\\
\tilde{\Phi}^{\prime}_{LM;\tau}(q) &=& \sum_{i=1}^A \left[ \left( {1 \over q} \overrightarrow{\nabla}_{{\bf{r}}_i} \times {\bf{M}}_{LL}^M(q {\bf{r}}_i) \right) \cdot \left(\vec{\sigma}_i \, \times {1 \over q} \overrightarrow{\nabla}_{{\bf{r}}_i} \right) + {1 \over 2} {\bf{M}}_{LL}^M(q {\bf{r}}_i) \cdot \vec{\sigma}_i \, \right]~t^\tau_{(i)} \nonumber \\
\Phi^{\prime \prime}_{LM;\tau}(q ) &=& i  \sum_{i=1}^A\left( {1 \over q} \overrightarrow{\nabla}_{{\bf{r}}_i}  M_{LM}(q {\bf{r}}_i) \right) \cdot \left(\vec{\sigma}_i \, \times \
{1 \over q} \overrightarrow{\nabla}_{{\bf{r}}_i}  \right)~t^\tau_{(i)} \,.
\label{eq:multipole}
\end{eqnarray}
where ${\bf{M}}_{LL}^{M}(q {\bf{r}}_i)=j_{L}(q r_i){\bf Y}^M_{LL1}(\Omega_{{\bf{r}}_i})$ , and $M_{LM}(q {\bf{r}}_i)=j_{L}(q r_i)Y_{LM}(\Omega_{{\bf{r}}_i})$. The vector spherical harmonics, ${\bf Y}^M_{LL1}(\Omega_{{\bf{r}}_i})$, are defined in terms of Clebsch-Gordan coefficients and scalar spherical harmonics:
\begin{equation}
{\bf Y}^M_{LL'1}(\Omega_{{\bf{r}}_i}) = \sum_{m\lambda} \langle L'm1\lambda|L'1LM \rangle
Y_{L'm}(\Omega_{{\bf{r}}_i}) \, {\bf e}_\lambda \,,
\end{equation}
where ${\bf e}_\lambda$ is a spherical unit vector basis. In Eq.~(\ref{eq:M}), we use the notation $W_{A}^{\tau \tau^\prime}(y)\equiv W_{AB}^{\tau \tau^\prime}(y)$ for $A=B$. 

The 6 nuclear response operators in~(\ref{eq:multipole}) arise from a multipole expansion of the nuclear charge and currents in Eq.~(\ref{eq:Hx}). The multipole index $L$ must be less than $2J$. In our analysis of current direct detection experiments, we adopt the nuclear response functions of Ref.~\cite{Anand:2013yka}. Nuclear response functions for dark matter capture by the Sun can be found in~\cite{Catena:2015uha} in analytic form.   
 
\begin{table}
    \centering
    \begin{tabular}{lcccc}
    \toprule
    Parameter         & Type & Prior range &  Prior type & Reference \\
    \midrule                                          
    $\log_{10} (c_1^\tau m_v^2)$ & model parameter & $[-5,1]$ & log-prior & - \\
    $\log_{10} (c_3^\tau m_v^2)$ & model parameter & $[-2,6]$ & log-prior & - \\
    $\log_{10} (c_4^\tau m_v^2)$ & model parameter & $[-3,3]$ & log-prior & - \\
    $\log_{10} (c_5^\tau m_v^2)$ & model parameter & $[-2,6]$ & log-prior & - \\
    $\log_{10} (c_6^\tau m_v^2)$ & model parameter & $[-2,6]$ & log-prior & - \\
    $\log_{10} (c_7^\tau m_v^2)$ & model parameter & $[-2,6]$ & log-prior & - \\
    $\log_{10} (c_8^\tau m_v^2)$ & model parameter & $[-2,6]$ & log-prior & - \\
    $\log_{10} (c_9^\tau m_v^2)$ & model parameter & $[-2,6]$ & log-prior & - \\
    $\log_{10} (c_{10}^\tau m_v^2)$ & model parameter & $[-2,6]$ & log-prior & - \\
    $\log_{10} (c_{11}^\tau m_v^2)$ & model parameter & $[-4,4]$ & log-prior & - \\
    $\log_{10} (c_{12}^\tau m_v^2)$ & model parameter & $[-2,6]$ & log-prior & - \\
    $\log_{10} (c_{13}^\tau m_v^2)$ & model parameter & $[-2,6]$ & log-prior & - \\
    $\log_{10} (c_{14}^\tau m_v^2)$ & model parameter & $[-2,6]$ & log-prior & - \\
    $\log_{10} (c_{15}^\tau m_v^2)$ & model parameter & $[-1,7]$ & log-prior & - \\
    $\log_{10} (m_{\chi}/{\rm GeV})$ & model parameter & $[0.5,4]$  & log-prior & - \\
     \midrule  
     $\xi_{\rm Xe}$ & nuisance &   $[0.78,0.86]$ & Gaussian. $\sigma=0.04$ & \cite{Arina:2011si} \\	
     $a_{\rm COUPP}$ & nuisance &   $[0.13,0.17]$ & Gaussian. $\sigma=0.02$ & \cite{Behnke:2012ys} \\	
     $a_{\rm PICASSO}$ & nuisance &   $[2.5,7.5]$ & Gaussian. $\sigma=2.50$ & \cite{Archambault:2012pm} \\	
    \bottomrule
    \end{tabular}
    \caption{List of model and nuisance parameters with corresponding prior types and ranges. The nuisance parameters $\xi_{\rm Xe}$, $a_{\rm COUPP}$, and $a_{\rm PICASSO}$ are introduced in Sec.~\ref{sec:datasets} to model various types of detector uncertainties. Following~\cite{Anand:2013yka}, we have expressed the coupling constants in units of $m_v^{2}~=~(246.2~{\rm GeV})^{2}$. }
\label{tab:prior}
\end{table}

\subsection{Scattering rate}
In a dark matter direct detection experiment, the expected differential rate of scattering events per unit time and per unit detector mass is given by
\begin{equation}
\frac{{\rm d}\mathcal{R}}{{\rm d}E_{R}} =  \sum_{T}\frac{{\rm d}\mathcal{R}_{T}}{{\rm d}E_{R}} \equiv  \sum_{T} \xi_T \frac{\rho_{\chi}}{2\pi m_\chi}  
 \int_{v > v_{\rm min}(q)} \,  \frac{f(\vec{v} + \vec{v}_e(t))}{v} \, \langle |\mathcal{M}_{NR}|^2\rangle_{\rm spins} \, d^3v 
\label{rate_theory}
\end{equation}
where $m_\chi$ is the dark matter particle mass, $\xi_T$ is the mass fraction of the nucleus $T$ in the target material, and $\rho_\chi$ is the dark matter density in the solar neighborhood. In Eq.~(\ref{rate_theory}) the integral denotes an average over the local dark matter velocity distribution, $f$, in the galactic rest frame boosted to the detector frame, 
$v_{\rm min}(q)=q/2\mu_T$ is the minimum velocity required to transfer a momentum $q$ from the target nucleus to the dark matter particle, and $\vec{v}_e(t)$ is the time-dependent Earth velocity in the galactic rest frame. Here we consider a Maxwell-Boltzmann distribution $f(\vec{v} + \vec{v}_e(t))\propto \exp(-|\vec{v}+\vec{v}_{e}(t)|^2/v_0^2)$ truncated at the local escape velocity $v_{\rm esc}=554$ km~s$^{-1}$, and with $v_{0}=220$ km~s$^{-1}$.

\section{Statistical framework}
\label{sec:statistics}
We compare the effective theory of dark matter-nucleon isoscalar and isovector interactions reviewed in Sec.~\ref{sec:EFT} to current dark matter direct detection experiments in a global multidimensional statistical analysis. 
If not otherwise specified, for a given dataset $\mathbf{d}$ we assume the Likelihood function
\begin{equation}
-\ln \mathcal{L}(\mathbf{d}|m_\chi,\mathbf{c},\bfeta,\mu_B) = \mu_S(m_\chi,\mathbf{c},\bfeta)+\mu_B - k - k \ln \left[\frac{\mu_S(m_\chi,\mathbf{c},\bfeta)+\mu_B}{k}\right] \,.
\label{Like}
\end{equation}    
In Eq.~(\ref{Like}) $\mu_S(m_\chi,\mathbf{c},\bfeta)$ is the number of predicted scattering events at a given point in parameter space, $k$ is the number of observed recoils in a given experiment, and $\mu_B$ is the corresponding number of expected background events. The arrays $\mathbf{c}$ and $\bfeta$ represent the 28 coupling constants of the theory and the nuisance parameters in Tab.~\ref{tab:prior}, respectively.
When the error $\sigma_B$ on the number of expected background events $\mu_B$ is known, we analytically marginalize Eq.~(\ref{Like}) over $\mu_B$ obtaining in the limit $\sigma_B \ll \mu_B$
\begin{eqnarray}
-\ln \mathcal{L}_{\rm eff}(\mathbf{d}|m_\chi,\mathbf{c},\bfeta) 
&\simeq& \mu_S(m_\chi,\mathbf{c},\bfeta) + (2-k) \ln \Big[\mu_S(m_\chi,\mathbf{c},\bfeta)+\mu_B\Big]  \nonumber\\
&-&\ln \left\{ \frac{(k^2-k)}{2}\sigma_B^2 + \left[ \mu_S(m_\chi,\mathbf{c},\bfeta)+\mu_B - \frac{k}{2}\sigma_B^2 \right]^2 \right\} \nonumber\\
&+& {\rm constant} \,.
\label{Like_eff}
\end{eqnarray}    
The constant term in the last line of Eq.~(\ref{Like_eff}) is an arbitrary normalization, that we choose such that $\ln \mathcal{L}_{\rm eff}=0$ for $\mu_S+\mu_B=k$.

We construct approximate 2D frequentist confidence intervals for pairs of model parameters from an effective chi-square defined as $\Delta \chi^2_{\rm eff}\equiv-2 \ln \mathcal{L}_{\rm prof}/ \mathcal{L}_{\rm max}$, where $\mathcal{L}_{\rm max}$ is the maximum Likelihood, and $\mathcal{L}_{\rm prof}$ is the 2D profile Likelihood:
\begin{equation}
\mathcal{L}_{\rm prof}(\mathbf{d}|\theta_1,\theta_2) \propto \max_{\theta_3,\dots,\theta_m} \mathcal{L}_{\rm eff}(\mathbf{d}|\mathbf{\Theta}) \,.
\label{eq:prof_Likelihood}
\end{equation} 
In Eq.~(\ref{eq:prof_Likelihood}), $\mathbf{\Theta}=(\theta_1,\theta_2,\dots,\theta_m)\equiv(m_\chi,\mathbf{c},\bfeta)$. Wilks' theorem guarantees that under certain regularity conditions the distribution of $\Delta \chi^2_{\rm eff}$ converges to a chi-square distribution with 2 degrees of freedom~\cite{2011JHEP...06..042F}.
  
Here we mainly restrict the analysis to a frequentist approach in order to avoid the volume effects found in~\cite{Catena:2014uqa} studying a smaller sample of dark matter-nucleon interaction operators. However, in computing exclusion limits that do not involve marginalization, we also adopt a Bayesian approach, since it requires a smaller number of Likelihood evaluations. Within the Bayesian approach, we express exclusion limits on the coupling constants in terms of x\% credible regions. Credible regions are defined as portions of the parameter space containing x\% of the total posterior probability, and such that the posterior probability density function $\mathcal{P}(\mathbf{\Theta}|\mathbf{d}) \propto  \mathcal{L}_{\rm eff}(\mathbf{d}|\mathbf{\Theta}) \pi(\mathbf{\Theta})$ at any point $\mathbf{\Theta}$ inside the credible region is larger than at any external point. 

The function $\pi(\mathbf{\Theta})$ is the prior probability density function. We assume a uniform prior probability density function for the logarithm of $m_\chi$ and $c_i^\tau$, with $\tau=0,1$ and $i=1,3,\dots,15$. This choice allows us to sample the Likelihood function covering several orders of magnitude in all directions in parameter space. For the nuisance parameters we assume Gaussian priors, as shown in Tab.~\ref{tab:prior}. 
  
We sample the multidimensional Likelihood function (\ref{Like_eff}) using the {\sffamily Multinest} program~\cite{Feroz:2008xx,Feroz:2007kg,Feroz:2013hea}. We run {\sffamily Multinest} with parameters set to $n_{\rm live}=20000$ and ${\rm tol}=10^{-4}$. 
Finally, we use our own routines to evaluate scattering rates and the Likelihood function, {\sffamily Getplots}~\cite{Austri:2006pe} to calculate the profile Likelihood, and {\sffamily Matlab} to produce the figures. 

\section{Datasets} 
\label{sec:exp}

\label{sec:datasets}
Here we list the experiments considered in the analysis. For each experiment we provide information needed to evaluate the corresponding Likelihood function. We refer to~\cite{Catena:2014uqa} for further details. In the analysis, we convolve Eq.~(\ref{rate_theory}) with a Gaussian probability distribution function of standard deviation $\sigma$ to account for finite energy resolution effects.

\subsection{CDMS-Ge}
\label{CDMS-Ge}
We consider data from the final exposure of the CDMS II experiment, corresponding to an exposure of 612 kg-days~\cite{Ahmed:2009zw}. In the data analysis the CDMS collaboration has observed $k=2$ candidate events within the signal region 10 - 100 keV, with an expected number of background events in the same energy interval equal to $\mu_B=0.9\pm0.3$.
Here we assume a maximum experimental efficiency of 32\% at 20 keV, linearly decreasing to 20\% at 10 keV and at 100 keV. In addition, we consider a Gaussian energy resolution with standard deviation $\sigma=0.2$. 

\subsection{CDMS low threshold}
\label{CDMS-LT}
In the 2 keV threshold analysis performed by the CDMS collaboration, the T1Z5 germanium detector has observed $k=36$ events in the 2 - 20 keV region~\cite{Ahmed:2010wy}. Zero-charge, surface, and bulk events, can explain 75\% of the observed counts~\cite{Ahmed:2010wy}. In computing the Likelihood function, we therefore consider $\mu_B=36\times0.75$, with $\sigma_{B}=0$. We assume the efficiency reported in the inset of Fig.~1 in~\cite{Ahmed:2010wy}, a Gaussian energy resolution with energy dependent standard deviation $\sigma=\sqrt{0.293^2+(0.056 E_R)^2}$, and an exposure for T1Z5 of 241/8 kg-days.

\subsection{SuperCDMS}
The SuperCDMS experiment has collected data using 7 germanium detectors, with a total exposure of 577 kg-days~\cite{Agnese:2014aze}. Tab.~1 in~\cite{Agnese:2014aze} provides details regarding the number of recorded events, as well as the number of expected background events for each detector separately. In our analysis we consider data from 5 detectors only, hence neglecting T5Z2 and T5Z3, since for these detectors the number of observed counts is significantly larger than the number of expected background events. In the calculations we consider a Gaussian energy resolution with $\sigma=0.3$, and the detector efficiency of Fig.~1 in Ref.~\cite{Agnese:2014aze}. We assume an average exposure per detector of 577/7 kg-days.
In the Likelihood function we set $\mu_B$ and $\sigma_B$ as in Tab.~1 of Ref.~\cite{Agnese:2014aze}.

\subsection{CDMSlite}
The SuperCDMS experiment has also collected data in the low ionization threshold operating mode (i.e. CDMSlite) during 10 live days of dark matter search~\cite{Agnese:2013jaa}. The nuclear recoil energy threshold for this experimental configuration is 170 eV$_{\rm ee}$, corresponding to 841 eV$_{\rm nr}$. The average rate of nuclear recoils in the CDMSlite detector is $5.2 \pm 1$ counts/keV$_{\rm ee}$/kg-day between 0.2 and 1 keV$_{\rm ee}$, and $2.9 \pm 0.3$ counts/keV$_{\rm ee}$/kg-day between 2 and 7 keV$_{\rm ee}$. For these data we assume the Likelihood function 
\begin{eqnarray}
-\ln \mathcal{L}_{\rm CDMSlite} &=& \frac{1}{2}\Theta_{\rm H}(\mathcal{R}_{[0.2,1]}-5.2)(\mathcal{R}_{[0.2,1]}-5.2)^2 \nonumber\\
&+&  \frac{1}{2}\Theta_{\rm H}(\mathcal{R}_{[2,7]}-2.9)(\mathcal{R}_{[2,7]}-2.9)^2/0.3^2 
\end{eqnarray}
where $\mathcal{R}_{[0.2,1]}$ and $\mathcal{R}_{[2,7]}$ represent the average rates between 0.2 and 1 keV$_{\rm ee}$, and between 2 and 7 keV$_{\rm ee}$, respectively. 
We calculate the expected average count rate in the CDMSlite detector using the efficiency reported in the inset of Fig.~1 in~\cite{Agnese:2013jaa}, and assuming perfect energy resolution. 

\subsection{XENON100}
\label{XENON100}
We use the data presented by the XENON100 collaboration in~\cite{Aprile:2012nq}, corresponding to an effective exposure of 34$\times$224.6 kg-days. 
XENON100 has observed $k=2$ candidate signal events in the pre-defined nuclear recoil energy range 6.6 - 30.5 keV$_{\rm nr}$, corresponding to 3 - 30 photoelectrons. The number of expected background events in the same interval is 1.0 $\pm$ 0.2. 
The differential spectrum of expected photoelectrons S1 in XENON100 is given by 
\begin{equation}
\frac{{\rm d}\mathcal{R}}{{\rm d}S1} = \mathcal{E}(S1)\sum_{n=1}^{+\infty} {\rm Gauss}(S1 | n,\sqrt{n}\sigma_{\rm PMT}) \,
\int_{0}^{\infty}{\rm d}E_{R} \,{\rm Poiss}(n|\nu(E_{R})) \frac{{\rm d}\mathcal{R}}{{\rm d}E_{R}} 
\label{eq:xenon100}
\end{equation}
In Eq.~(\ref{eq:xenon100}), $n$ is the number of photoelectrons actually produced from a recoil energy $E_R$. For $n$ we assume a Poisson distribution of mean $\nu(E_{R})=E_{R} L_{\rm eff}(E_{R}) L_y S_{\rm nr}/S_{\rm ee}$, where $L_y = 2.28 \pm 0.04$ PE/keV$_{\rm ee}$ is the light yield, $S_{\rm ee} = 0.58$ and $S_{\rm nr} = 0.95$ are the electric field scintillation quenching factors for electron and nuclear recoils, and $L_{\rm eff}(E_{R})$ is the energy dependent scintillation efficiency. We assume for $L_{\rm eff}(E_{R})$ the following form
\begin{equation}
L_{\rm eff}(E_{R})  = \left\{
\begin{array}{ll}
\bar{L}_{\rm eff}(E_{R}) \qquad  \qquad  \qquad  \qquad  \qquad  \qquad  \qquad  \,\quad \,\,\,{\rm for} \,\, E_{R}/{\rm keV}_{\rm nr} \ge 3 \nonumber\\
\nonumber\\
\max\{\xi_{\rm Xe}[\ln(E_{R}/{\rm keV}_{\rm nr}) -\ln3]+0.09, 0\} \qquad {\rm for} \,\,1 < E_{R}/{\rm keV}_{\rm nr} < 3
\end{array} \right.
\end{equation}
where $\bar{L}_{\rm eff}(E_{R})$ is the best-fit scintillation efficiency reported in Fig.~1 of Ref.~\cite{Aprile:2011hi}, and $\xi_{\rm Xe}$ is a nuisance parameter, first proposed in Ref.~\cite{Arina:2011si} to logarithmically extrapolate the scintillation efficiency below 3 keV. In Eq.~(\ref{eq:xenon100}) we also assume a Gaussian photomultiplier resolution of variance $n\sigma^2_{\rm PMT}$, with $\sigma_{\rm PMT}=0.5$. Finally we adopt the detector efficiency $\mathcal{E}(S_1)$ in Fig.~2 of~\cite{Aprile:2011hi}.

\subsection{XENON10}
\label{XENON10}
In a second study the XENON collaboration uses the ionization signal S2 only to measure the nuclear recoil energy of the detected events~\cite{Angle:2011th}. The experimental recoil energy threshold for this analysis is of about 1.4 keV$_{\rm nr}$. The relation between S2 (measured in PE) and the observed nuclear recoil energy is $S2=\mathcal{Q}_y(E_R) E_R$. We extract the function $\mathcal{Q}_y(E_R)$ from Fig.~1 in~\cite{Angle:2011th}. The experimental exposure corresponding to these data is 12.5$\times$1.2 kg-days. In this low energy threshold analysis XENON10 observed $k=23$ candidate events in the signal region 1.4 - 10 keV$_{\rm nr}$. A large background contamination cannot be excluded~\cite{Angle:2011th}. We calculate the expected nuclear recoil energy spectrum in XENON10 as in Eq.~5.5 of~\cite{Lewin:1995rx}, assuming a constant efficiency $ \mathcal{E}(E_R)=0.94$. We also allow for $\mu_B=20$ background events of unknown origin, with $\sigma_B=10$. 

\subsection{LUX}
\label{LUX}
We construct the LUX Likelihood function as for XENON100. The first data release of the LUX experiment consists of 85.3 live days of dark matter search data, corresponding to an exposure of 250$\times$85.3 kg-days~\cite{Akerib:2013tjd}. LUX observed 160 events with S1 between 2 and 30 PE. Only one event is below the mean of the Gaussian fit to the nuclear recoil calibration events in Fig.~4 of~\cite{Akerib:2013tjd}. The expected number of background events in the same region is $0.64\pm0.16$. We assume $\sigma_{\rm PMT}=0.37$, and the experimental efficiency in Fig.~9 of Ref.~\cite{Akerib:2013tjd} divided by 2 to account for the 50\% nuclear recoil acceptance quoted by the LUX collaboration.

\subsection{COUPP}
\label{COUPP}
Using a 4.0 kg CF$_3$I bubble chamber, the COUPP experiment has measured $k=2$,  $k=3$ and $k=8$ bubble nucleations above the threshold energies 7.8 keV$_{\rm nr}$, 11 keV$_{\rm nr}$ and 15.5 keV$_{\rm nr}$, respectively~\cite{Behnke:2012ys}. We calculate the expected number of dark matter scattering events above a threshold energy $E_{\rm th}$ as follows~\cite{Behnke:2012ys} 
\begin{equation}
\mu_{S}(m_\chi,\mathbf{c},\bfeta) =\epsilon(E_{\rm th}) \sum_{T={\rm C,F,I}} \int_{E_{\rm th}}^{\infty}{\rm d}E_{R} \, \mathcal{P}_T(E_{R},E_{\rm th}) \frac{{\rm d}\mathcal{R}_{T}}{{\rm d} E_{R}}
\label{coupp}
\end{equation}
where $\epsilon(E_{\rm th})$ is the threshold dependent experimental exposure times the bubble detection efficiency, and $\mathcal{P}_{T}(E_{R},E_{\rm th})$ is the probability that an energy $E_{R}$ nucleates a bubble above $E_{\rm th}$. The latter is given by~\cite{Behnke:2012ys}
\begin{equation}
\mathcal{P}_{T}(E_{R},E_{\rm th}) = 1 - \exp\left[ -\alpha_{T} \frac{E_{R}-E_{\rm th}}{E_{\rm th}}\right] \,.
\end{equation}
Here we neglect scattering from Carbon ($\alpha_{\rm C}=0$), assume perfect efficiency for bubble nucleation for Iodine ($\alpha_{\rm I}\rightarrow +\infty$), and treat  $\alpha_{\rm F}\equiv a_{\rm COUPP}$ as a nuisance parameter. For the energy dependent exposure we assume $\epsilon(7.8~{\rm keV})=55.8$ kg-days, $\epsilon(11~{\rm keV})=70$ kg-days and $\epsilon(15.5~{\rm keV})=311.7$ kg-days~\cite{Behnke:2012ys}. For the three threshold configurations the expected number of background events is $\mu_B=0.8$, $\mu_B=0.7$ and $\mu_B=3$, respectively. 

\subsection{PICASSO}
\label{PICASSO}
The PICASSO experiment searches for dark matter using superheated liquid droplets made of C$_4$F$_{10}$~\cite{Archambault:2012pm}. In the last run PICASSO operated assuming eight different bubble nucleation threshold energies, namely 1.7, 2.9, 4.1, 5.8, 6.9, 16.3, 38.8 and 54.8 keV. For each energy, the observed count rate above threshold  $\hat{\mathcal{R}}_i$, $i=1,\dots,8$, is reported in Fig.~5 of Ref.~\cite{Archambault:2012pm}, including the associated errors $\sigma_i$. We calculate the expected scattering rate $\mathcal{R}_i$ in the energy range $[E_{\rm th},+\infty]$ using Eq.~(\ref{coupp}) with $\epsilon=1$. As for COUPP, we neglect scattering from Carbon, and consider $\alpha_{\rm F}=a_{\rm PICASSO}$ as a nuisance parameter. For the eight data points $\hat{\mathcal{R}}_i$ we assume the Gaussian Likelihood function:
\begin{equation}
-\ln \mathcal{L}_{\rm PICASSO}  = \sum_{i=1}^{8} \frac{1}{2\sigma_{i}^{2}} \left[  \mathcal{R}_i - \hat{\mathcal{R}}_i \right]^2 \,.
\end{equation}

\section{Global limits and interference patterns}
\label{sec:results}
We now compare the general effective theory of one-body dark matter-nucleon interactions to current observations, in a global statistical analysis of the direct detection experiments in Sec.~\ref{sec:exp}. We present our results in terms of exclusion limits on the 28 coupling constants of the theory. Our investigation focuses on multi-interaction interference effects in the exclusion limit calculation. 

Multi-interaction interference effects are of two types. A first type involves pairs of dark matter-nucleon interaction operators. 
It generates terms proportional to $c_i^{\tau}c_j^{\tau}$, with $i\neq j$, in Eq.~(\ref{rate_theory}), arising from $R_{\Sigma'}^{\tau\tau}$, $R_{\Sigma''}^{\tau\tau}$, $R_{\Phi'' M}^{\tau\tau}$ and $R_{\Delta\Sigma'}^{\tau\tau}$.
Inspection of Eq.~(\ref{eq:R}) in Appendix~\ref{sec:appDM} shows that there are 7 pairs of interfering operators in the general effective theory of one-body dark matter-nucleon interactions, namely $(\hat{\mathcal{O}}_1,\hat{\mathcal{O}}_3)$, $(\hat{\mathcal{O}}_4,\hat{\mathcal{O}}_5)$, $(\hat{\mathcal{O}}_4,\hat{\mathcal{O}}_6)$, $(\hat{\mathcal{O}}_8,\hat{\mathcal{O}}_9)$, $(\hat{\mathcal{O}}_{11},\hat{\mathcal{O}}_{12})$, $(\hat{\mathcal{O}}_{11},\hat{\mathcal{O}}_{15})$, and $(\hat{\mathcal{O}}_{12},\hat{\mathcal{O}}_{15})$. Other pairs of operators do not interfere, partially since nuclear ground states are eigenstates of $P$ and $CP$, and partially because of their $ {\bf{\hat{S}}}_\chi$ dependence. We refer to the remaining operators in Tab.~\ref{tab:operators} as non-interfering dark matter-nucleon interaction operators. 
\begin{table}
    \centering
    \begin{tabular}{ccc|ccc|ccc|ccc}
    \toprule
    $c_i^\tau$\,$c_j^{\tau'}$ & & $r_{ij}^{\tau\tau'}$ & $c_i^\tau$\,$c_j^{\tau'}$   & & $r_{ij}^{\tau\tau'}$ & $c_i^\tau$\,$c_j^{\tau'}$ & & $r_{ij}^{\tau\tau'}$ & $c_i^\tau$\,$c_j^{\tau'}$ & & $r_{ij}^{\tau\tau'}$ \\
    \midrule
   $c_1^0$\,$c_1^1$ && 1 &  $c_9^0$\,$c_9^1$ && 1 &  $c_1^0$\,$c_3^0$ && 0.89 & $c_1^1$\,$c_3^1$ && 0.49  \\
   $c_3^0$\,$c_3^1$  && 0.56 &  $c_{10}^0$\,$c_{10}^1$ && 1 &  $c_4^0$\,$c_5^0$ && -0.03 & $c_4^1$\,$c_5^1$ && -0.32  \\
   $c_4^0$\,$c_4^1$  && 1&  $c_{11}^0$\,$c_{11}^1$ && 1 & $c_4^0$\,$c_6^0$ && -0.64  & $c_4^1$\,$c_6^1$ &&  -0.65 \\
   $c_5^0$\,$c_5^1$  && 0.79 &  $c_{12}^0$\,$c_{12}^1$ && 0.57 &  $c_8^0$\,$c_9^0$ && 0.04 &  $c_8^1$\,$c_9^1$ &&  0.49 \\
   $c_6^0$\,$c_6^1$  && 1 &  $c_{13}^0$\,$c_{13}^1$ &&   1  & $c_{11}^0$\,$c_{12}^0$ && 1 & $c_{11}^1$\,$c_{12}^1$ && 0.55 \\
   $c_7^0$\,$c_7^1$  && 1&  $c_{14}^0$\,$c_{14}^1$ && 1&  $c_{11}^0$\,$c_{15}^0$ && -0.92  &  $c_{11}^1$\,$c_{15}^1$ && -0.49 \\
   $c_8^0$\,$c_8^1$  && 0.82 &  $c_{15}^0$\,$c_{15}^1$  && 0.55 &  $c_{12}^0$\,$c_{15}^0$&& 0.94& $c_{12}^1$\,$c_{15}^1$&& 0.93 \\
           \midrule
     \bottomrule
    \end{tabular}
    \caption{Table of correlation coefficients (\ref{eq:rij}) different from zero for the LUX experiment at $m_\chi=10$ TeV.}
    \label{tab:corr}
\end{table}

A second type of multi-interaction interference arises when isoscalar and isovector components of the same operator combine in Eq.~(\ref{rate_theory}) forming terms proportional to $c_i^{\tau}c_i^{\tau'}$, $\tau\neq \tau'$. Any operator in Tab.~\ref{tab:operators} can produce isoscalar-isovector interference patterns of this type as long as off-diagonal nuclear response functions are generated in the dark matter-nucleus scattering.  

Let us now consider a pair of operators $(\hat{\mathcal{O}}_i,\hat{\mathcal{O}}_j)$ and an experiment characterized by the signal region $\mathcal{S}$. Let us also assume $c_i^\tau\neq0$ and $c_j^{\tau'}\neq0$ for fixed $(\tau,\tau')$ and $(i,j)$ pairs, and all other coupling constants equal to zero. Then, contours of constant $\mu_S(m_\chi,\mathbf{c},\bfeta)$ at a given $m_\chi$ and $\bfeta$ represent ellipses in the $c_i^\tau-c_j^{\tau'}$ plane of equation
\begin{equation}
a_{ii}^{\tau\tau'} c_i^{\tau} c_i^{\tau'}  + 2 a^{\tau\tau'}_{ij} c_i^{\tau} c_j^{\tau'} + a^{\tau\tau'}_{jj} c_j^{\tau} c_j^{\tau'} = {\rm const}\,,
\label{eq:ellipse}
\end{equation}   
where the constants $a^{\tau\tau'}_{ii}$, $a^{\tau\tau'}_{jj}$, and $a^{\tau\tau'}_{ij}$ are given by
\begin{equation}
a^{\tau\tau'}_{ii}= \int_{\mathcal{S}} {\rm d}E_R ~ \frac{{\rm d}\mathcal{R}}{{\rm d}E_{R}} \Big|_{c_i^{\tau}=c_i^{\tau'}=1;\,\,c_j^{\tau}=c_j^{\tau'}=0}
\end{equation}
and similarly for $a^{\tau\tau'}_{jj}$ and $a^{\tau\tau'}_{ij}$. 

We can therefore introduce a correlation coefficient $r^{\tau\tau'}_{ij}$ to quantify the degree of interference between a pair of operators $(\hat{\mathcal{O}}_i,\hat{\mathcal{O}}_j)$, or between the isoscalar and isovector components of a given operator: 
\begin{equation}
r^{\tau\tau'}_{ij} = - \frac{a^{\tau\tau'}_{ij}}{\sqrt{a^{\tau\tau'}_{ii} a^{\tau\tau'}_{jj}}} \,.
\label{eq:rij}
\end{equation}
In Tab.~\ref{tab:corr} we list some of the $r^{\tau\tau'}_{ij}$ coefficients different from zero for the LUX experiment at $m_\chi=10$~TeV.

Below we fit $m_\chi$, $\boldsymbol{\eta}$, and ${\bf c}$ to observations {\it globally}, i.e. including all the available data, and simultaneously varying sets of coupling constants with $r_{ij}^{\tau\tau'}\neq0$. 

\begin{figure}[t]
\begin{center}
\begin{minipage}[t]{0.49\linewidth}
\centering
\includegraphics[width=\textwidth]{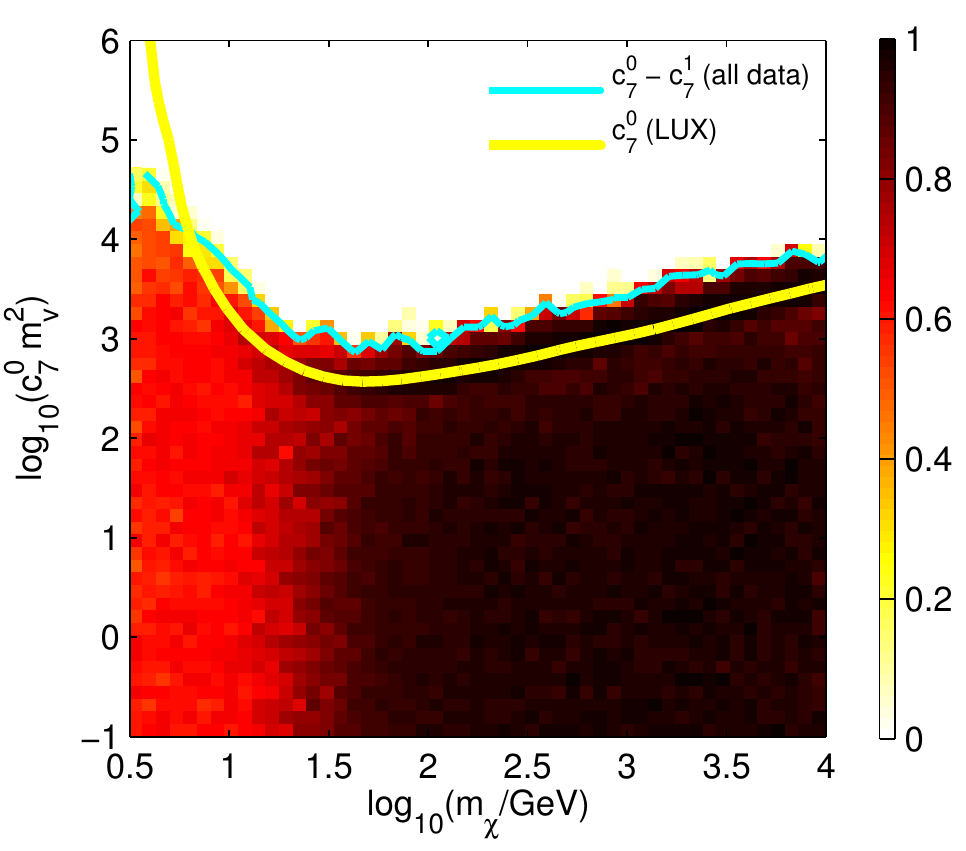}
\end{minipage}
\begin{minipage}[t]{0.49\linewidth}
\centering
\includegraphics[width=\textwidth]{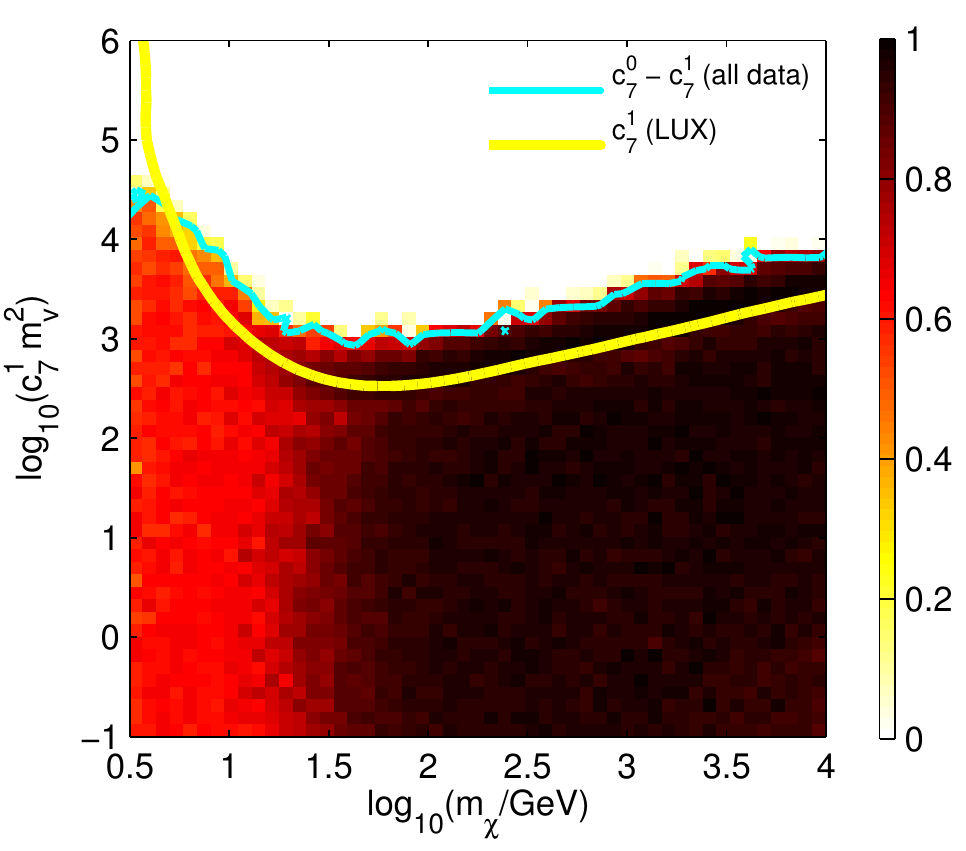}
\end{minipage}
\begin{minipage}[t]{0.49\linewidth}
\centering
\includegraphics[width=\textwidth]{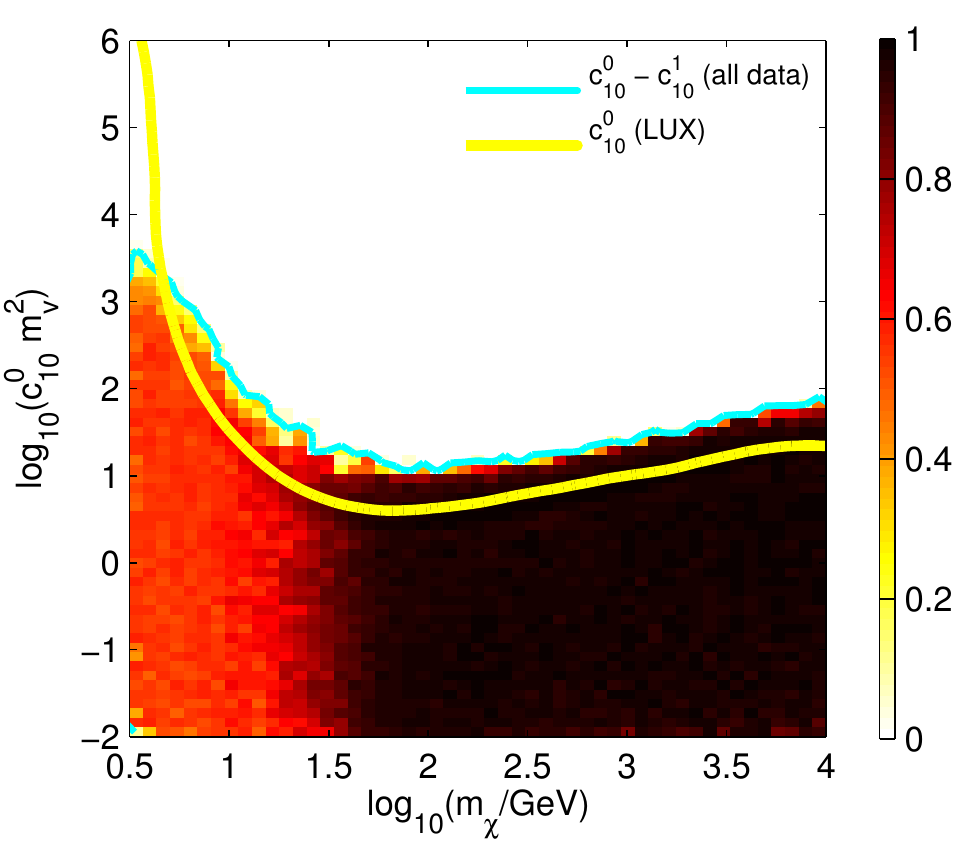}
\end{minipage}
\begin{minipage}[t]{0.49\linewidth}
\centering
\includegraphics[width=\textwidth]{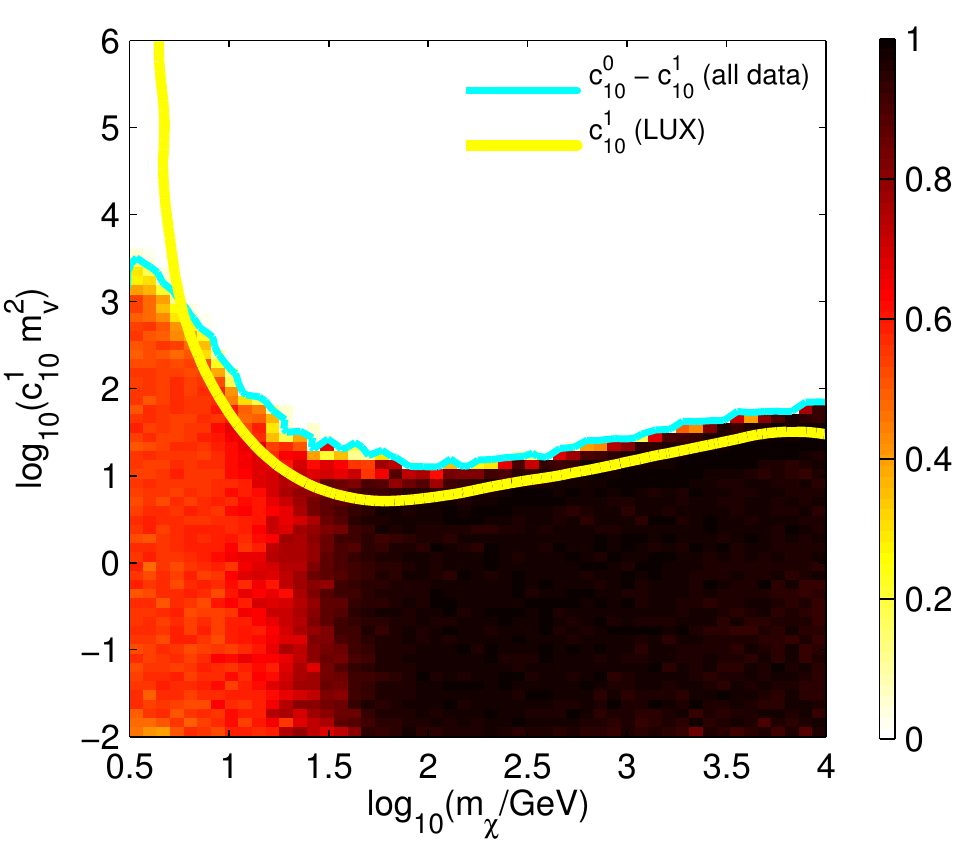}
\end{minipage}
\end{center}
\caption{In all panels, cyan contours represent 2D 90\% confidence intervals from a fit of $m_{\chi}$, $\boldsymbol{\eta}$ and of the coupling constants in the legends to the direct detection experiments indicated in parenthesis. In each panel, colored regions correspond to the 2D profile Likelihood associated with the cyan contours. Yellow contours denote 2D 90\% credible regions obtained by fitting $m_\chi$ and a single coupling constant to the LUX data.}
\label{fig:c7c10}
\end{figure}
\begin{figure}[t]
\begin{center}
\begin{minipage}[t]{0.49\linewidth}
\centering
\includegraphics[width=\textwidth]{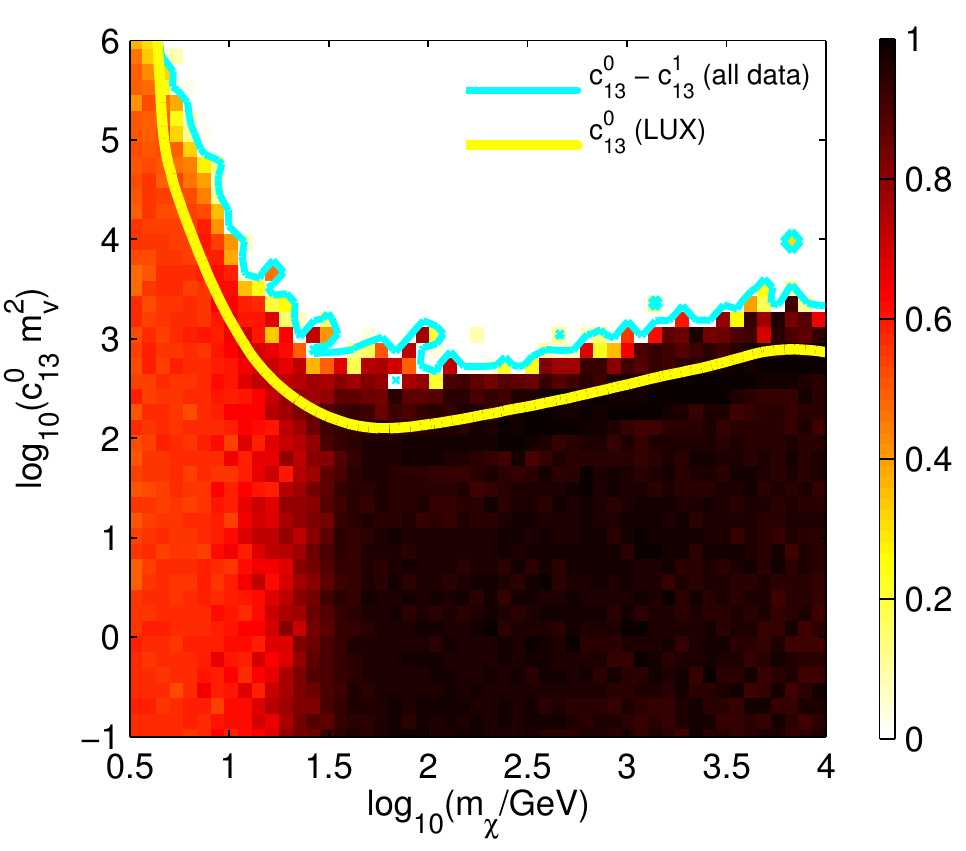}
\end{minipage}
\begin{minipage}[t]{0.49\linewidth}
\centering
\includegraphics[width=\textwidth]{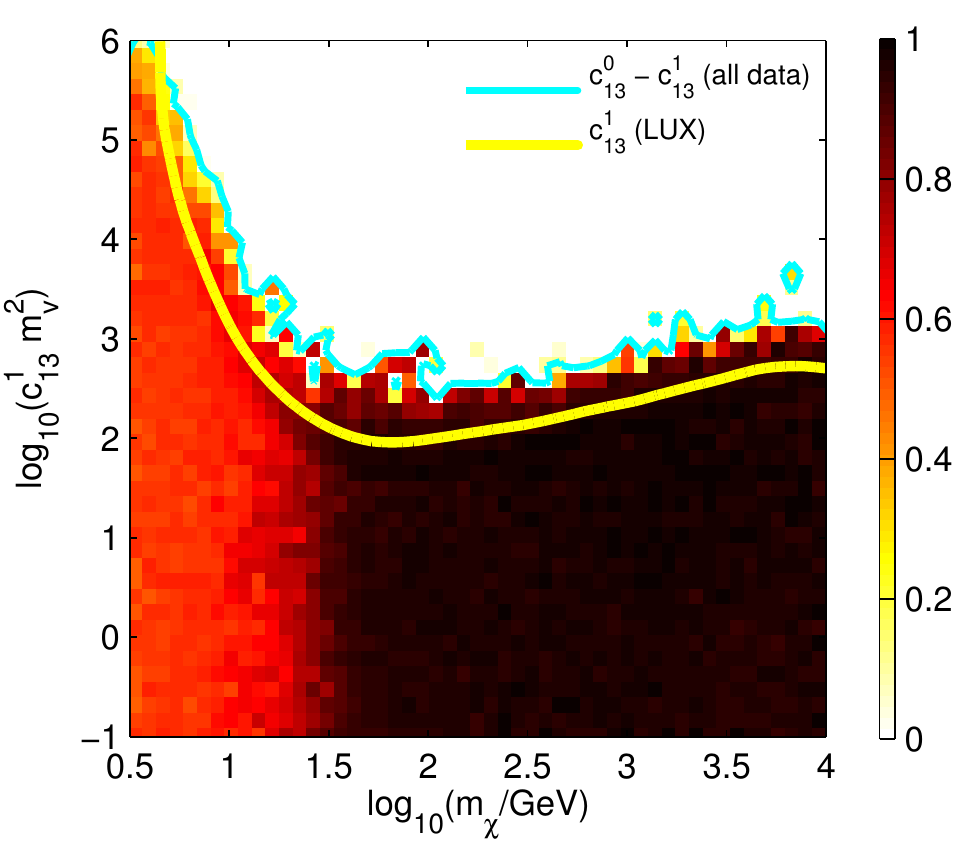}
\end{minipage}
\begin{minipage}[t]{0.49\linewidth}
\centering
\includegraphics[width=\textwidth]{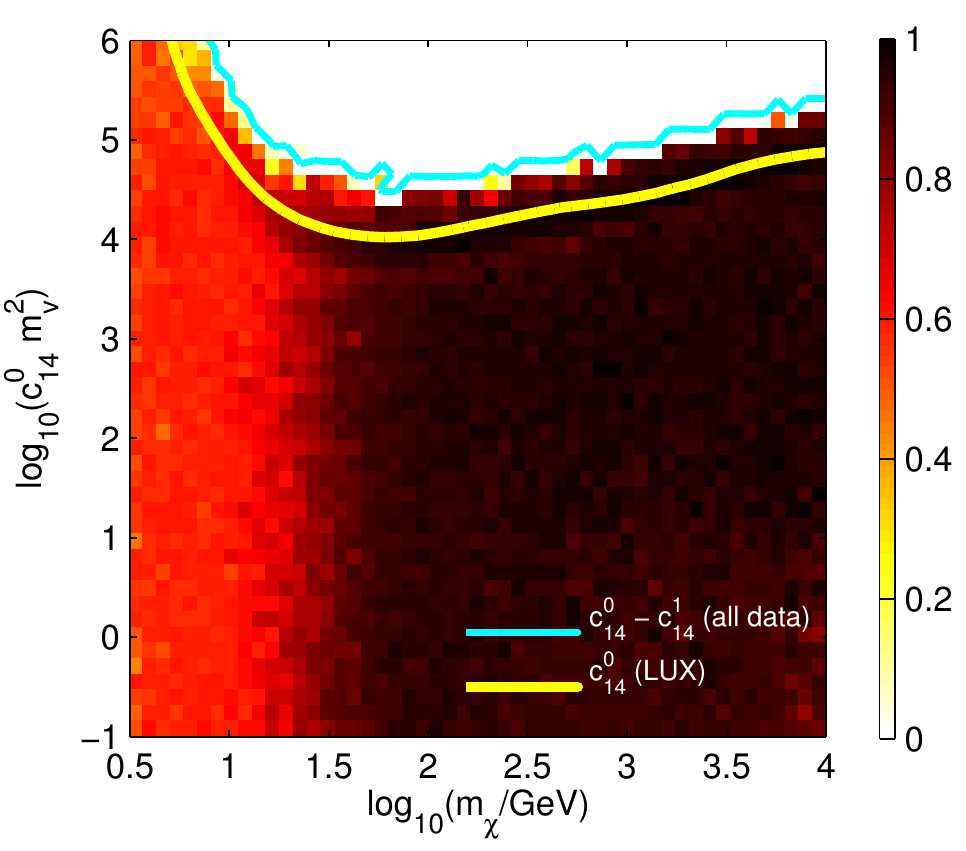}
\end{minipage}
\begin{minipage}[t]{0.49\linewidth}
\centering
\includegraphics[width=\textwidth]{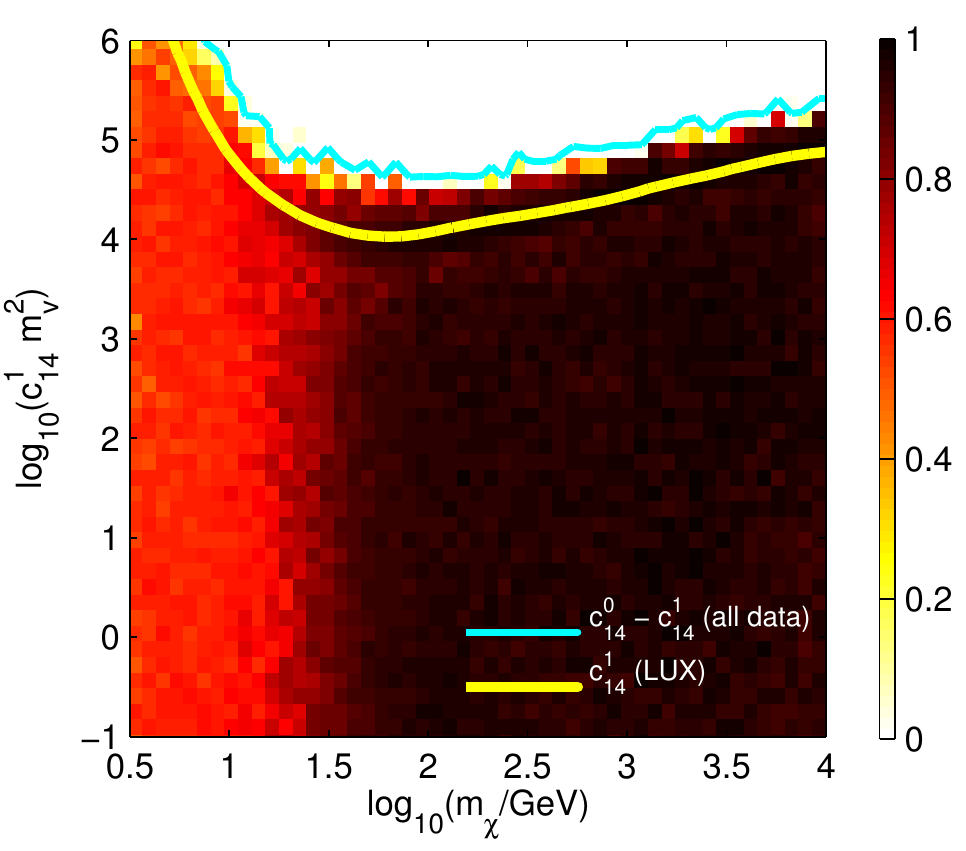}
\end{minipage}
\end{center}
\caption{Same as for Fig.~\ref{fig:c7c10} but now for the model parameters $c^{0}_{13}$, $c^{1}_{13}$, $c^{0}_{14}$, $c^{1}_{14}$ and $m_\chi$.}
\label{fig:c13c14}
\end{figure}
\begin{figure}[t]
\begin{center}
\begin{minipage}[t]{0.49\linewidth}
\centering
\includegraphics[width=\textwidth]{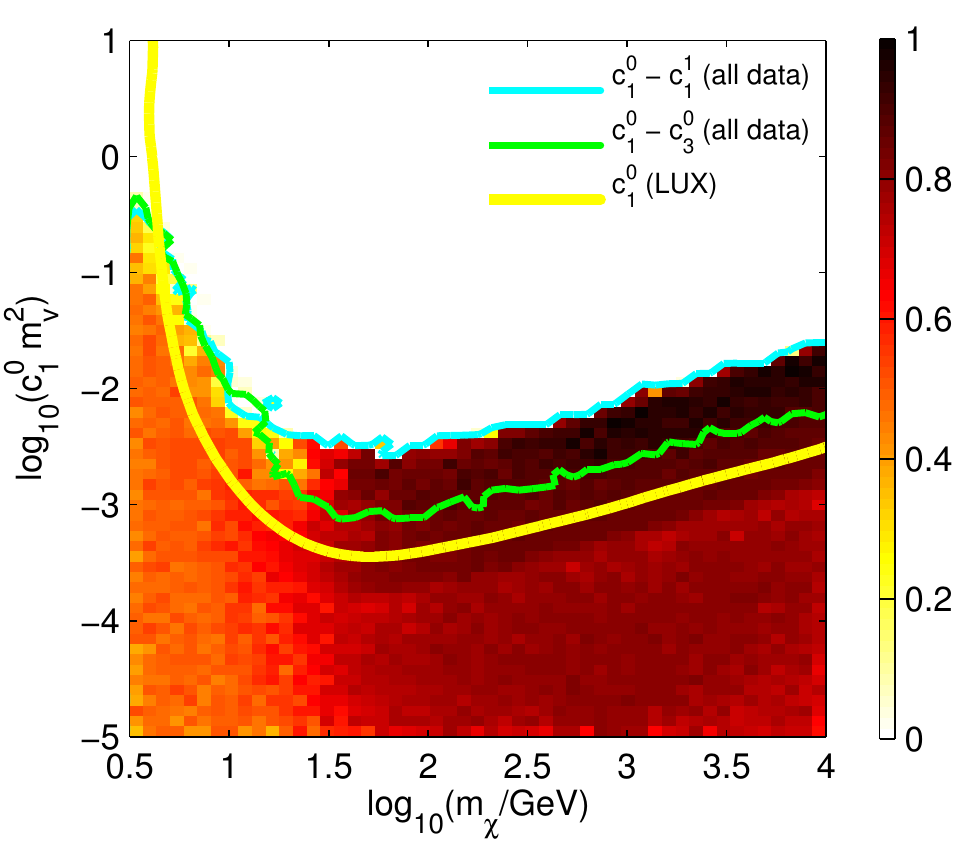}
\end{minipage}
\begin{minipage}[t]{0.49\linewidth}
\centering
\includegraphics[width=\textwidth]{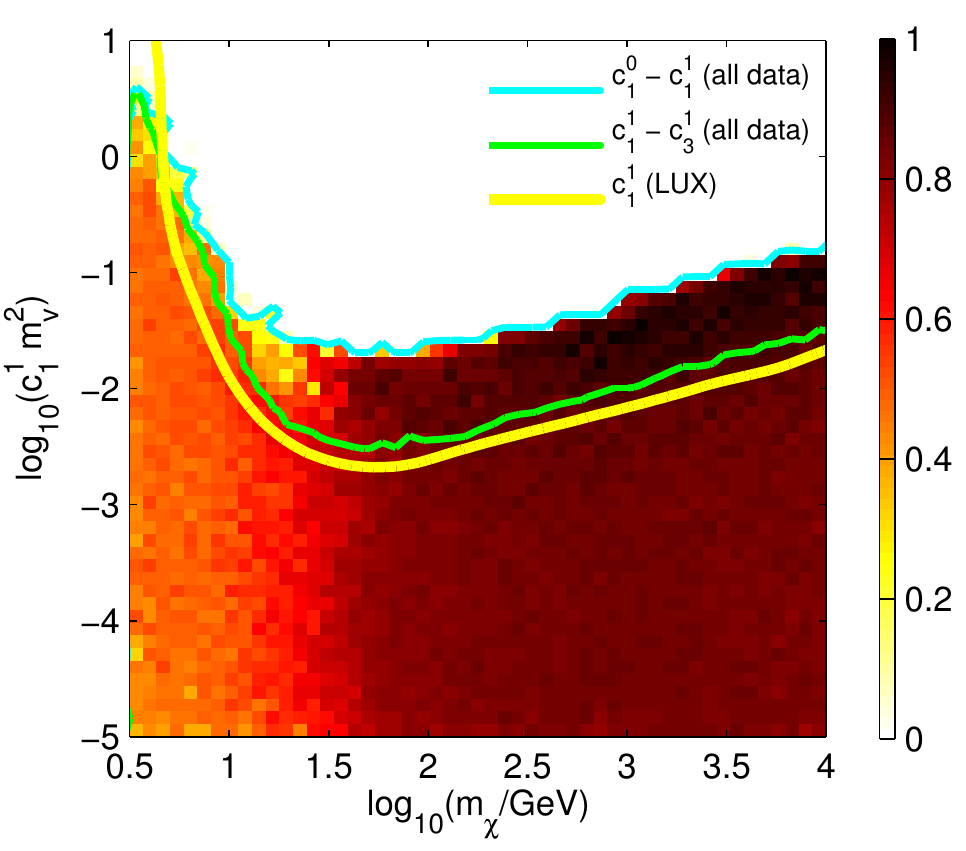}
\end{minipage}
\begin{minipage}[t]{0.49\linewidth}
\centering
\includegraphics[width=\textwidth]{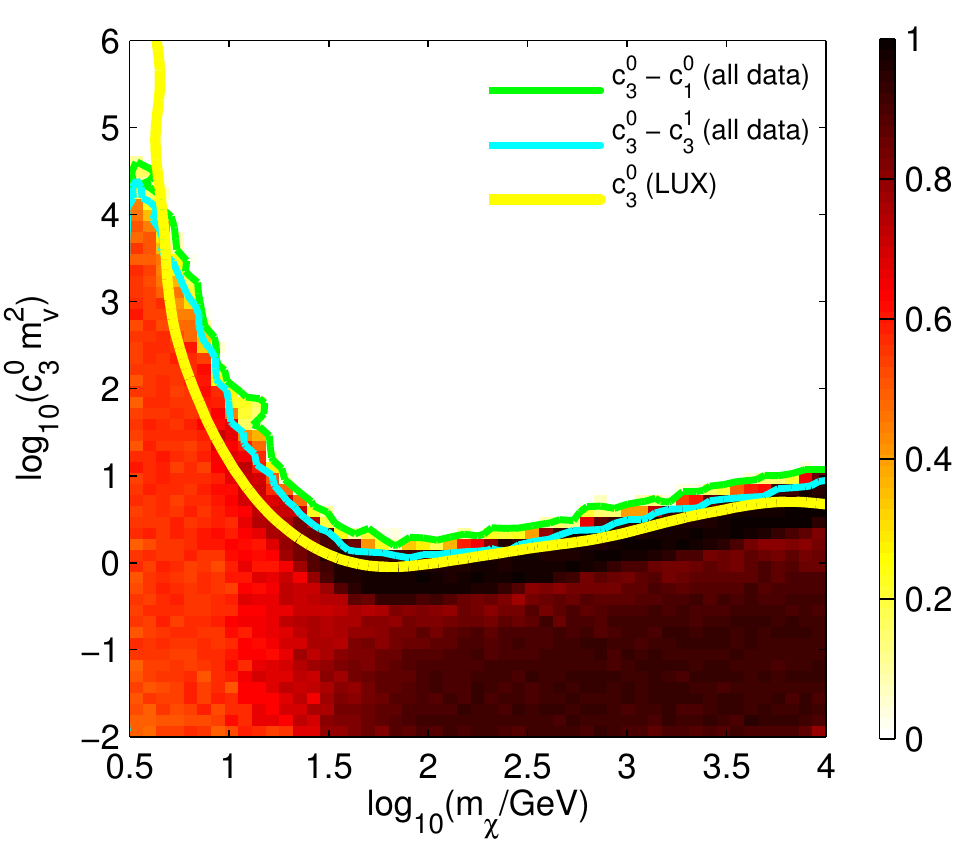}
\end{minipage}
\begin{minipage}[t]{0.49\linewidth}
\centering
\includegraphics[width=\textwidth]{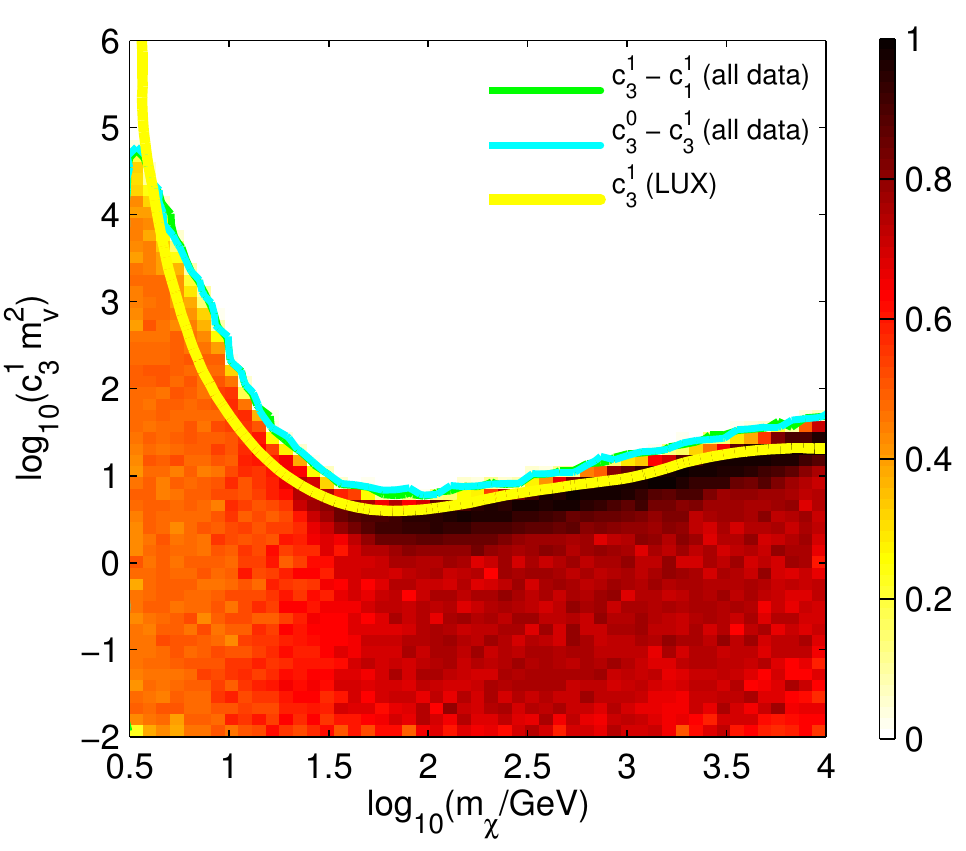}
\end{minipage}
\end{center}
\caption{In all panels, green and cyan contours represent 2D 90\% confidence intervals from a global fit of $m_\chi$, $\boldsymbol{\eta}$ and of the coupling constants in the legends to all direct detection experiments in Sec.~\ref{sec:exp}. In each panel, the colored region corresponds to the 2D profile Likelihood associated with the less stringent exclusion limit at the reference value $m_\chi\sim100$~GeV in that figure. Yellow contours represent 2D 90\% credible regions obtained by fitting $m_\chi$ and a single coupling constant to the LUX data.
}
\label{fig:c1c3}
\end{figure}

\begin{figure}[t]
\begin{center}
\begin{minipage}[t]{0.32\linewidth}
\centering
\includegraphics[width=\textwidth]{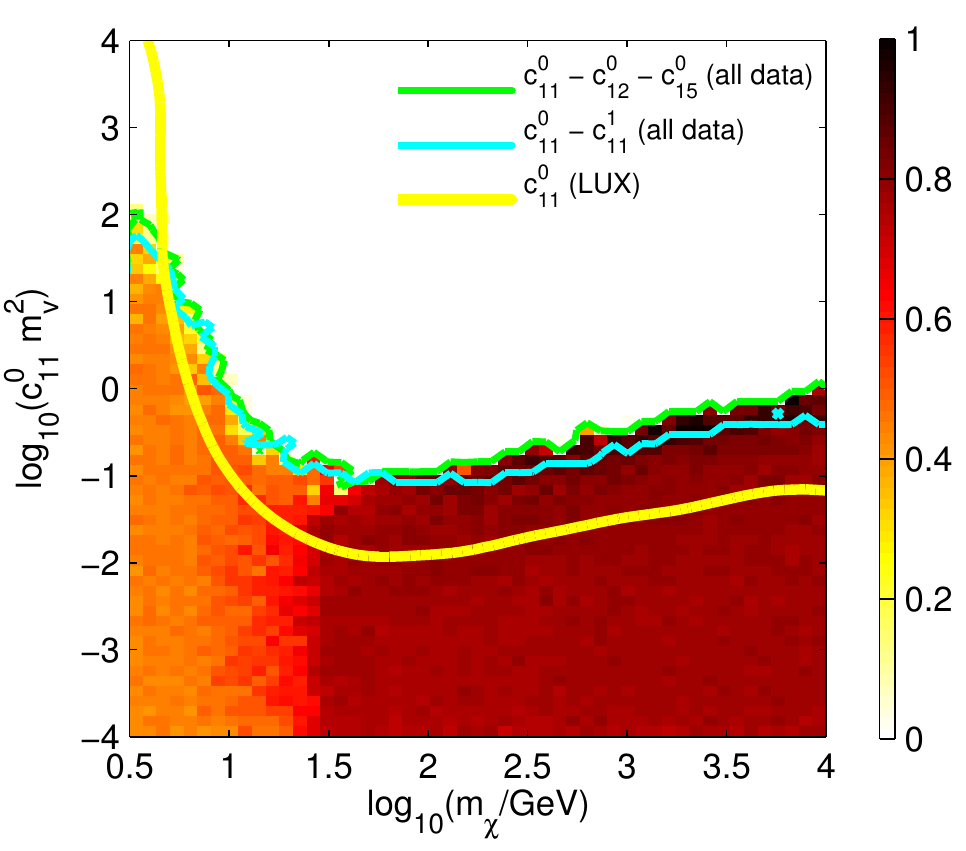}
\end{minipage}
\begin{minipage}[t]{0.32\linewidth}
\centering
\includegraphics[width=\textwidth]{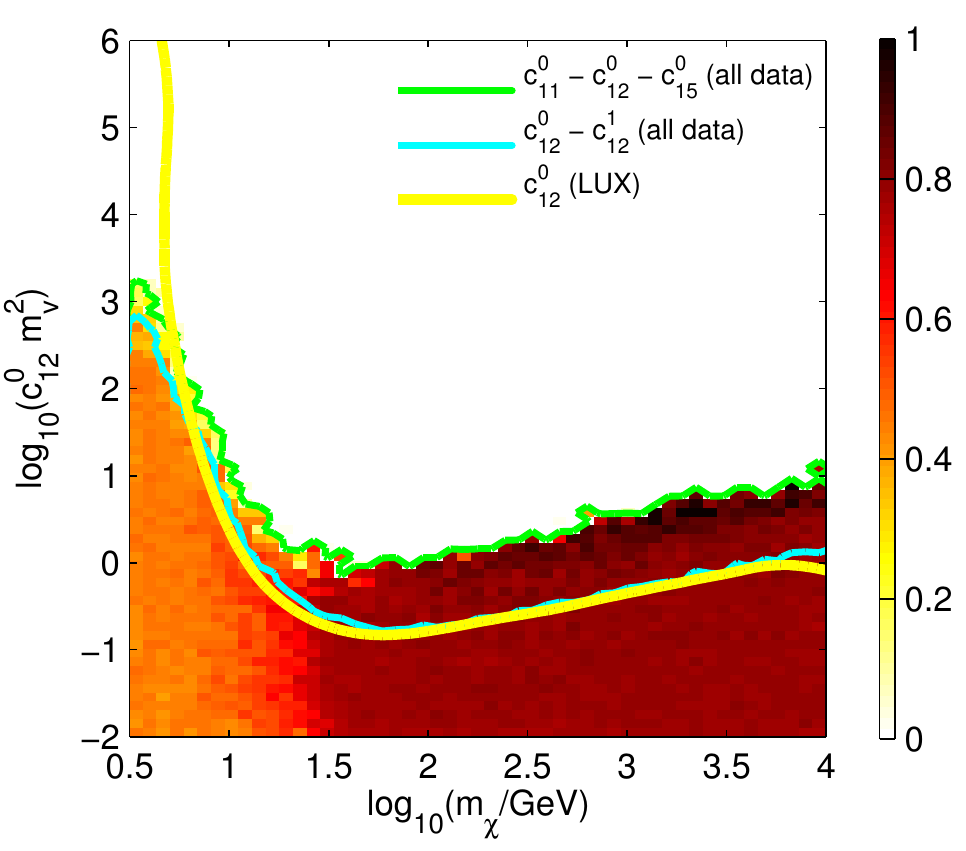}
\end{minipage}
\begin{minipage}[t]{0.32\linewidth}
\centering
\includegraphics[width=\textwidth]{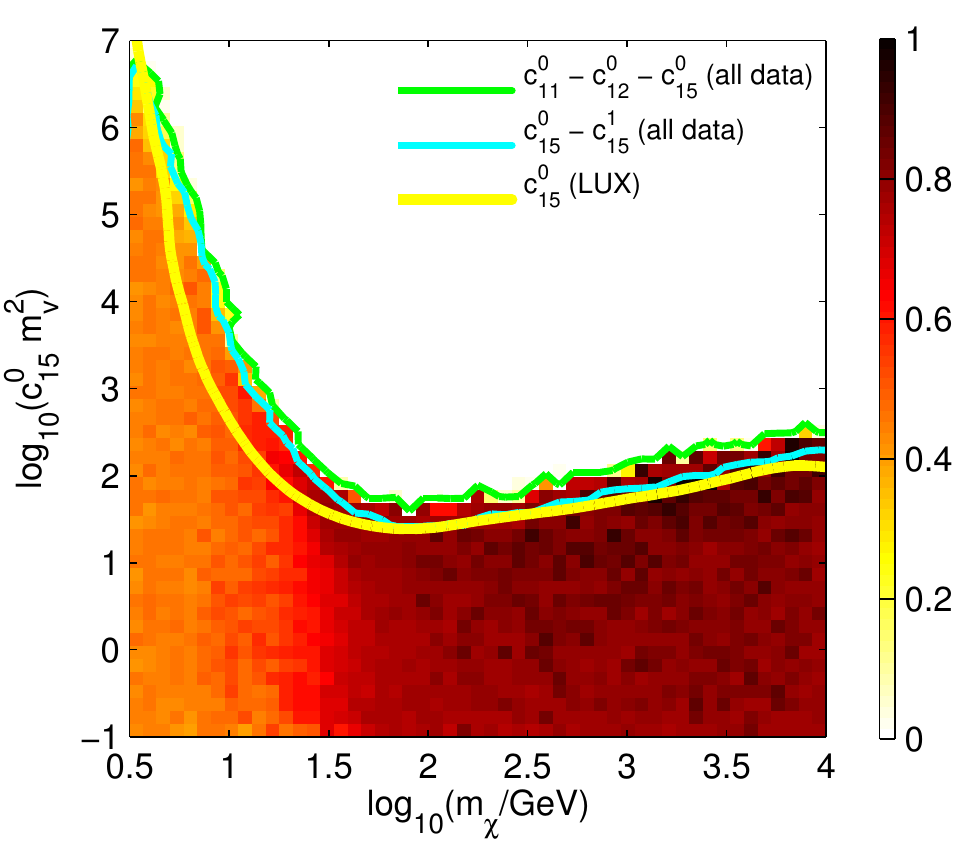}
\end{minipage}
\begin{minipage}[t]{0.32\linewidth}
\centering
\includegraphics[width=\textwidth]{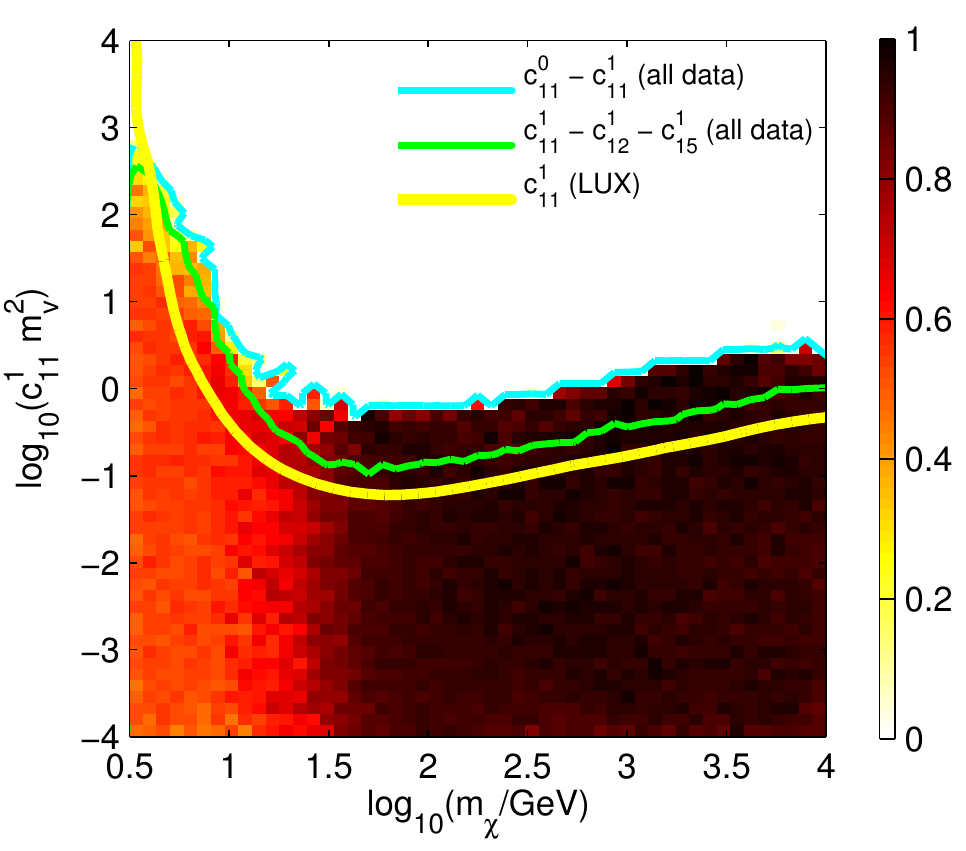}
\end{minipage}
\begin{minipage}[t]{0.32\linewidth}
\centering
\includegraphics[width=\textwidth]{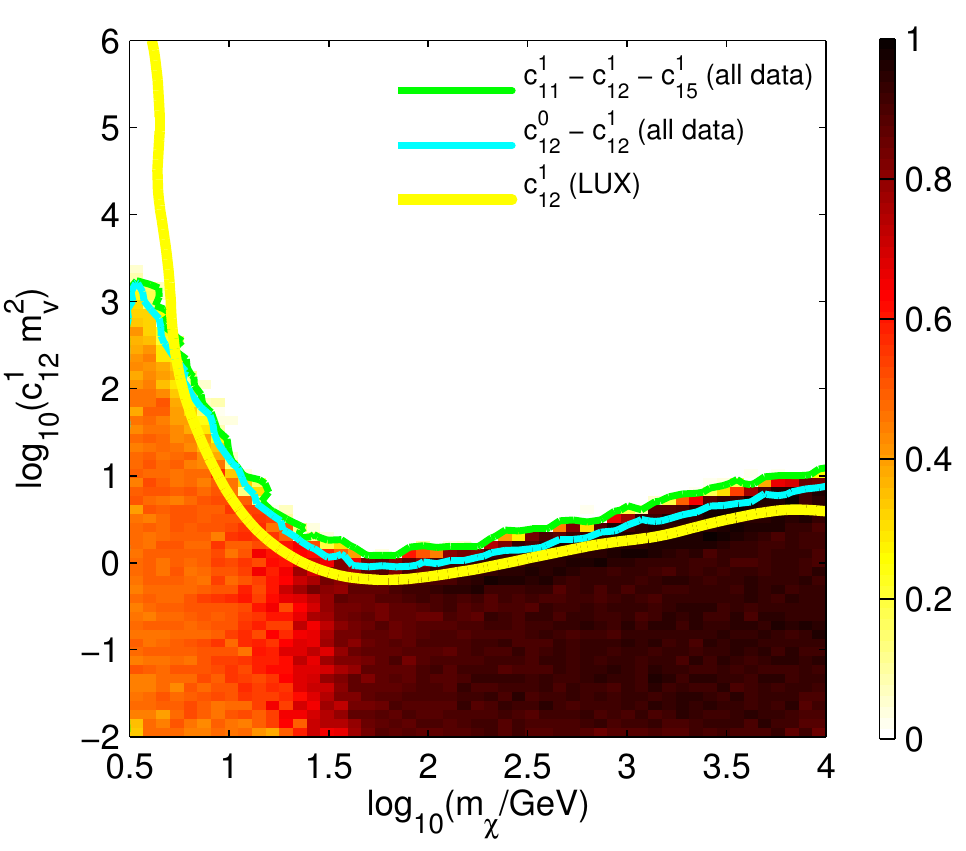}
\end{minipage}
\begin{minipage}[t]{0.32\linewidth}
\centering
\includegraphics[width=\textwidth]{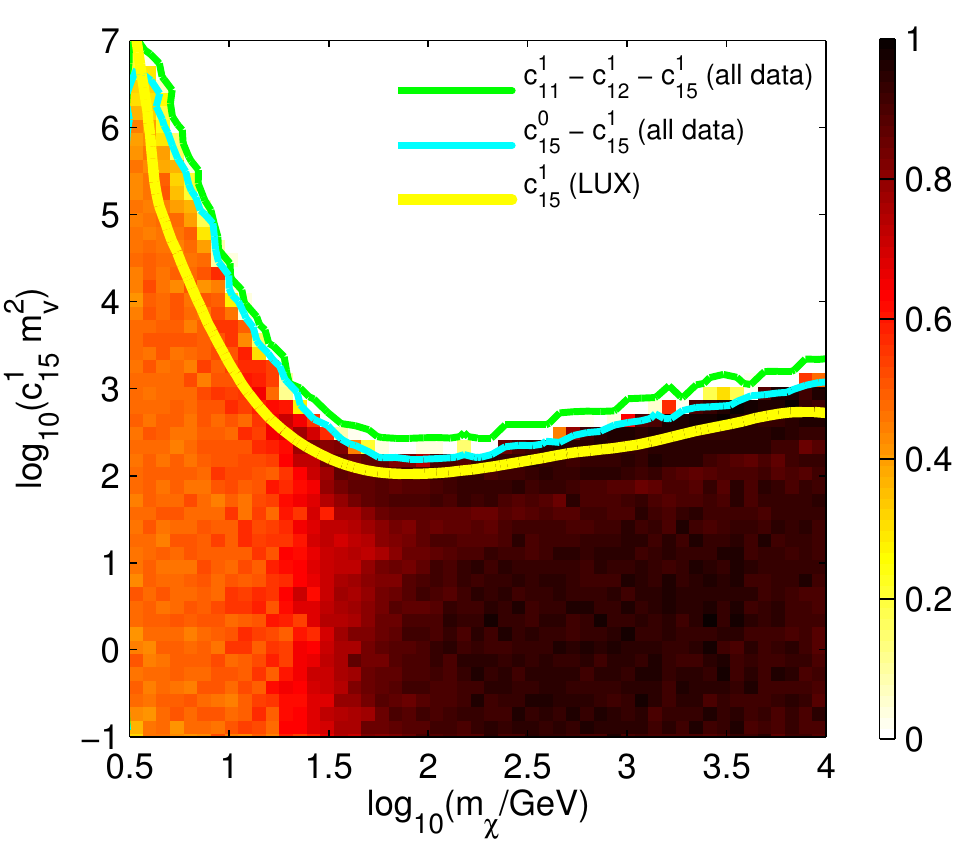}
\end{minipage}
\end{center}
\caption{Same as for Fig.~\ref{fig:c1c3}, but now for the operators $\hat{\mathcal{O}}_{11}$, $\hat{\mathcal{O}}_{12}$, and $\hat{\mathcal{O}}_{15}$.}
\label{fig:c11c12c15}
\end{figure}

\begin{figure}[t]
\begin{center}
\begin{minipage}[t]{0.32\linewidth}
\centering
\includegraphics[width=\textwidth]{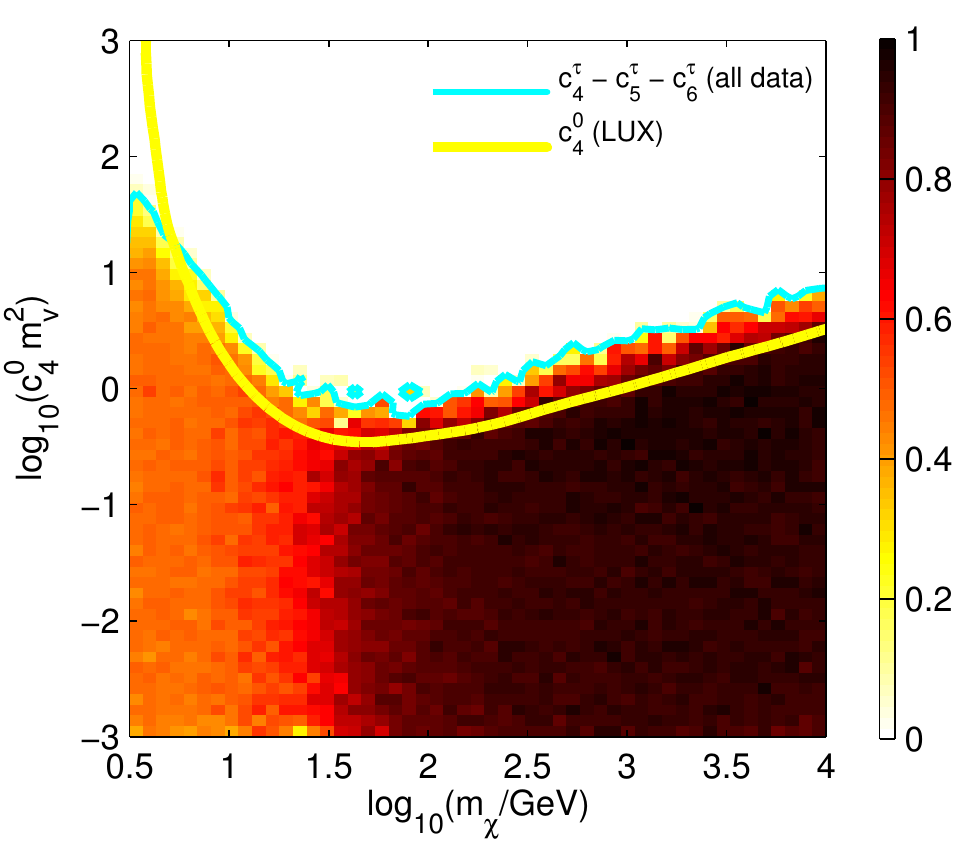}
\end{minipage}
\begin{minipage}[t]{0.32\linewidth}
\centering
\includegraphics[width=\textwidth]{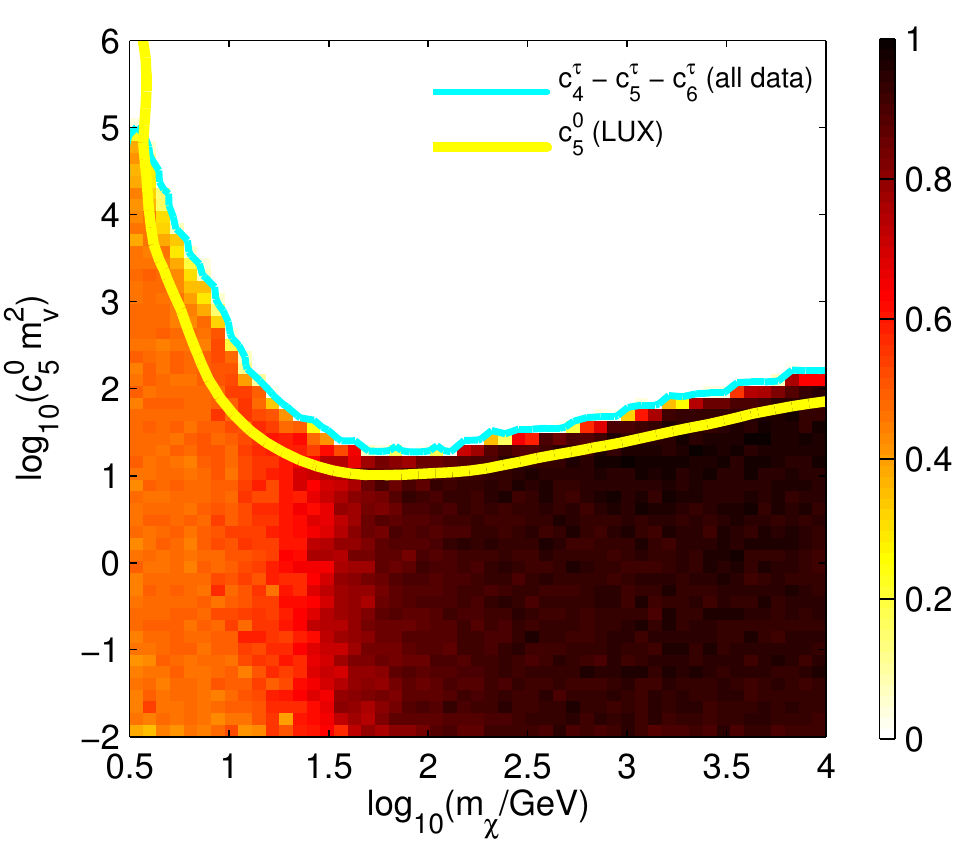}
\end{minipage}
\begin{minipage}[t]{0.32\linewidth}
\centering
\includegraphics[width=\textwidth]{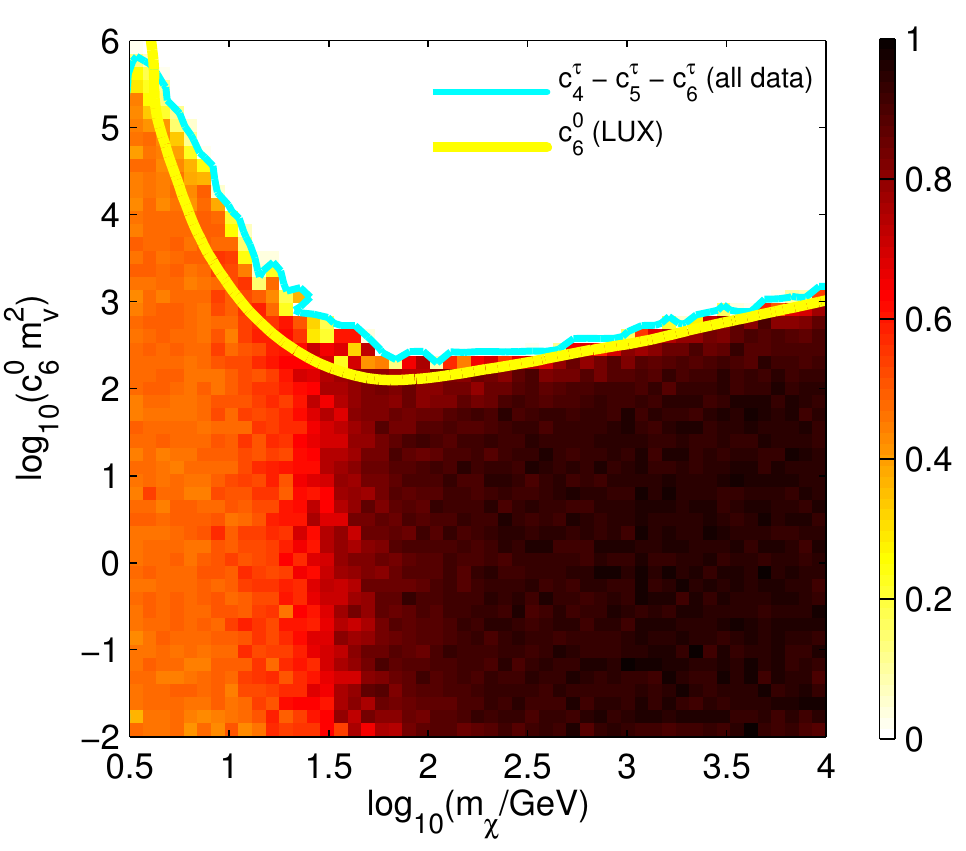}
\end{minipage}
\begin{minipage}[t]{0.32\linewidth}
\centering
\includegraphics[width=\textwidth]{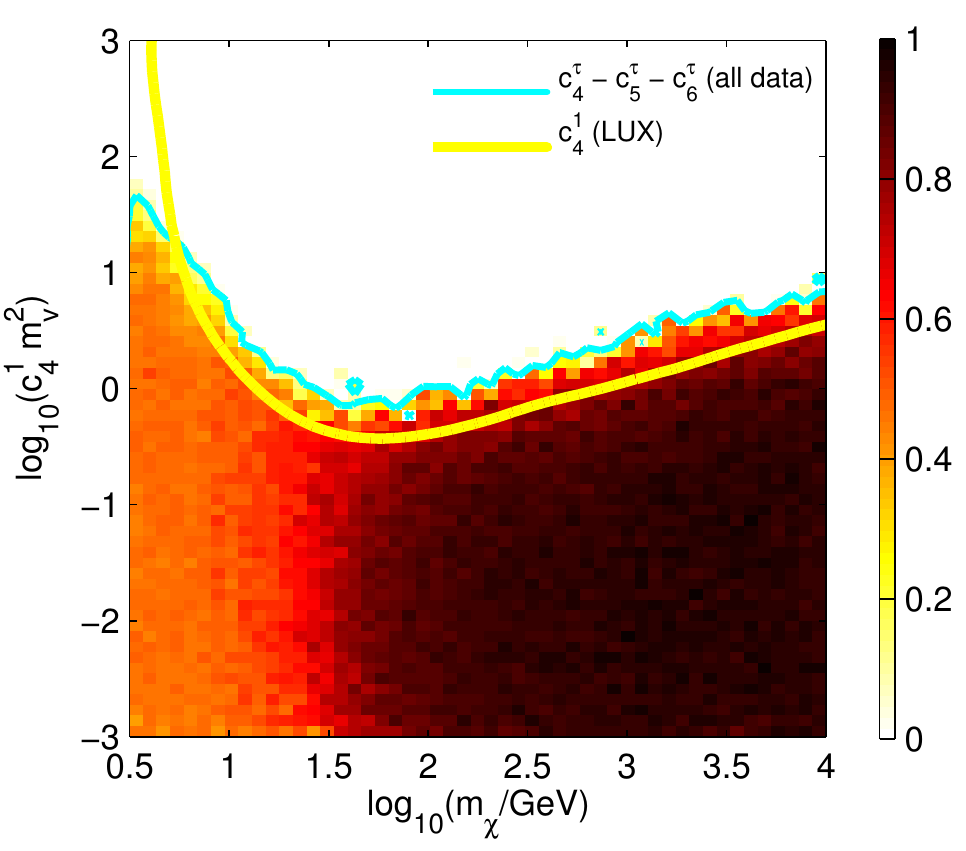}
\end{minipage}
\begin{minipage}[t]{0.32\linewidth}
\centering
\includegraphics[width=\textwidth]{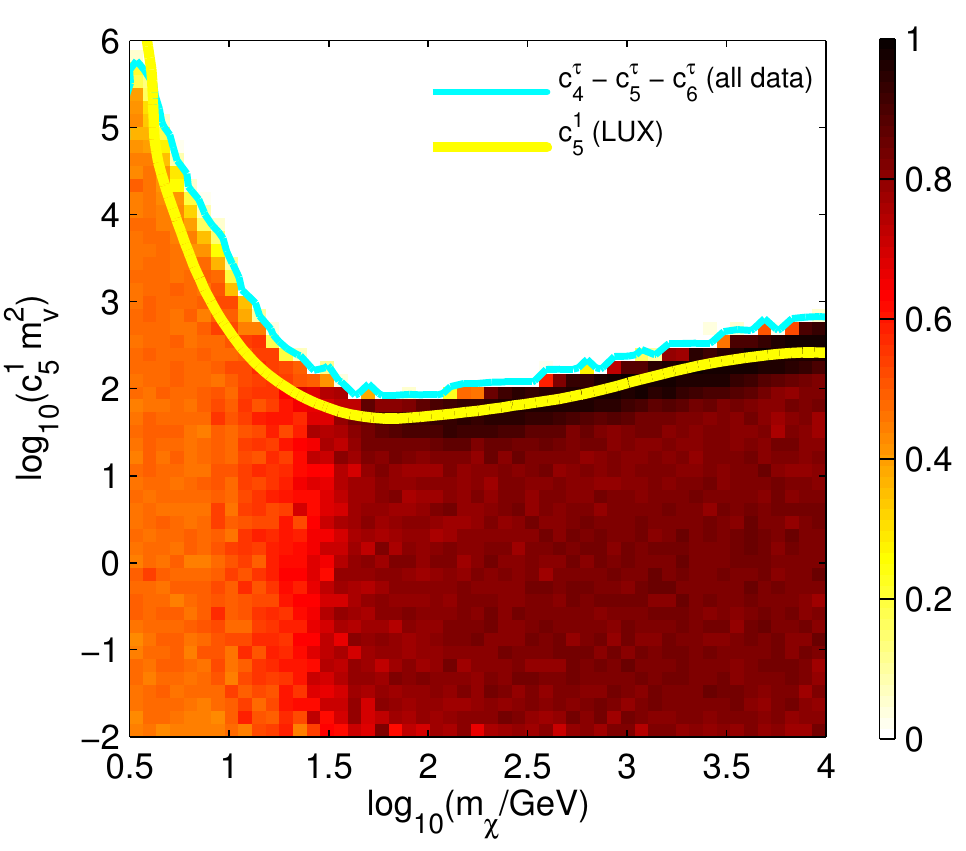}
\end{minipage}
\begin{minipage}[t]{0.32\linewidth}
\centering
\includegraphics[width=\textwidth]{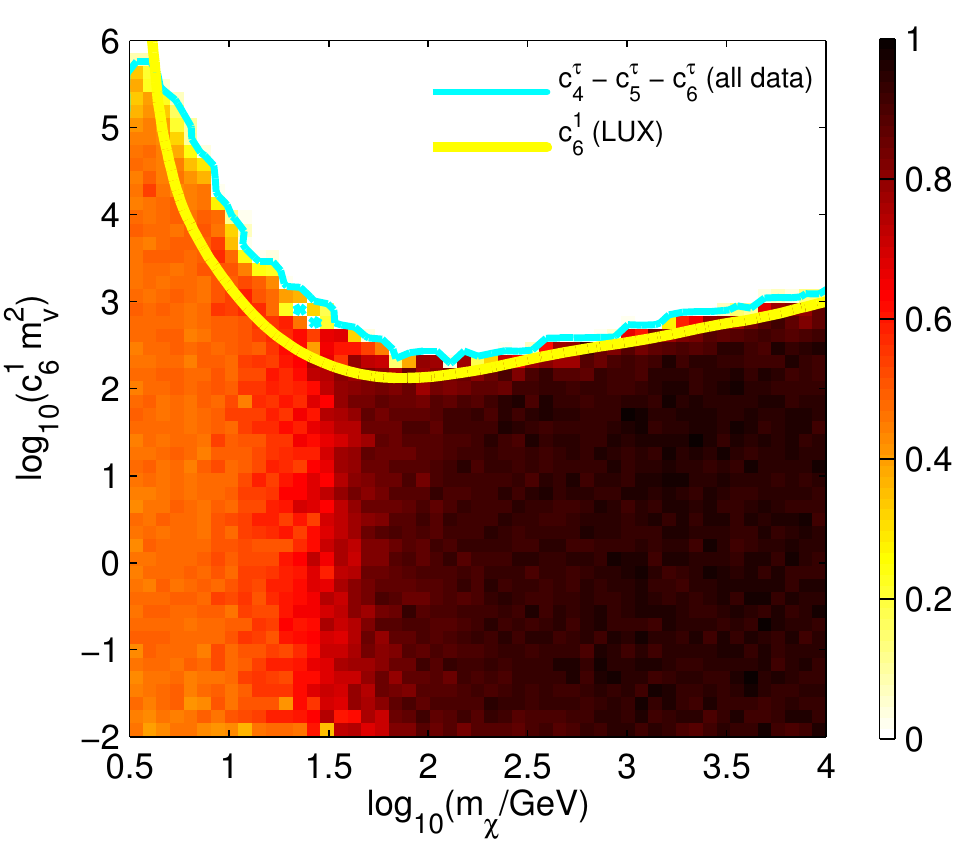}
\end{minipage}
\end{center}
\caption{2D 90\% confidence intervals (cyan contours) and profile Likelihoods (colored regions) from a global fit of the model parameters 
in the legends, $\boldsymbol{\eta}$ and $m_\chi$ to current direct detection experiments. Contours are presented in the six planes $c^{0}_{4}-m_\chi$, $c^{0}_{5}-m_\chi$, $c^{0}_{6}-m_\chi$, $c^{1}_{4}-m_\chi$, $c^{1}_{5}-m_\chi$, and $c^{1}_{6}-m_\chi$. Yellow lines represent 2D 90\% credible regions obtained by fitting $m_\chi$ and a single coupling constant to the LUX data.}
\label{fig:c4c5c6}
\end{figure}

\begin{figure}[t]
\begin{center}
\begin{minipage}[t]{0.49\linewidth}
\centering
\includegraphics[width=\textwidth]{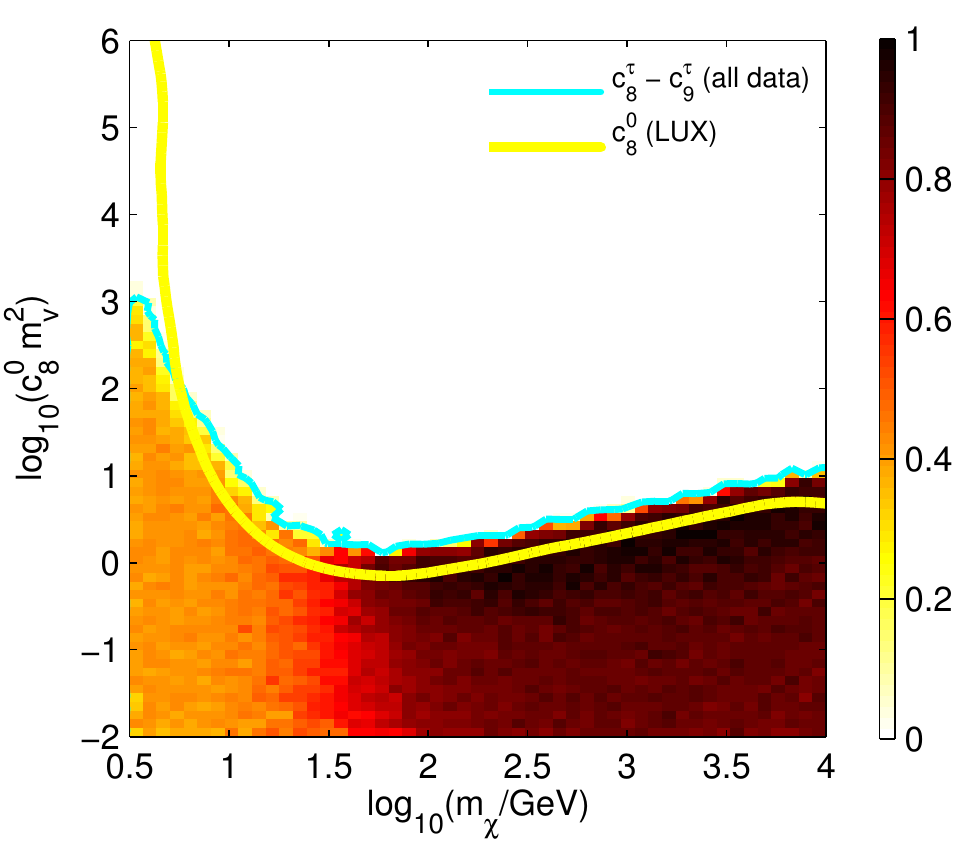}
\end{minipage}
\begin{minipage}[t]{0.49\linewidth}
\centering
\includegraphics[width=\textwidth]{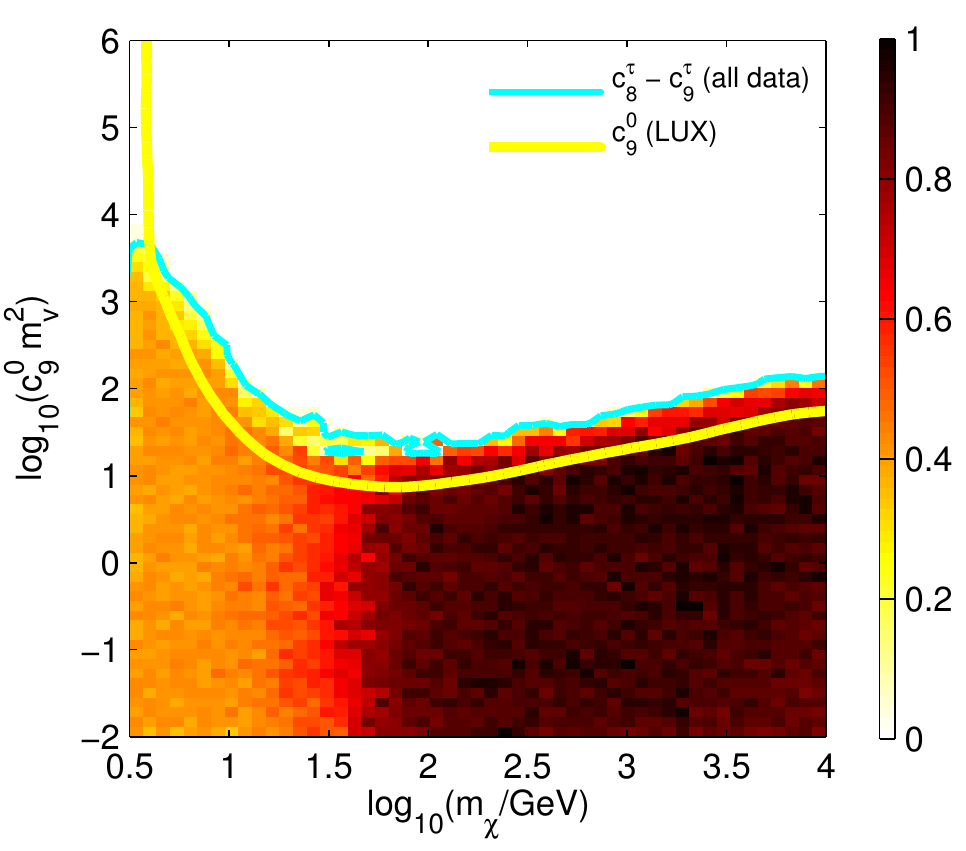}
\end{minipage}
\begin{minipage}[t]{0.49\linewidth}
\centering
\includegraphics[width=\textwidth]{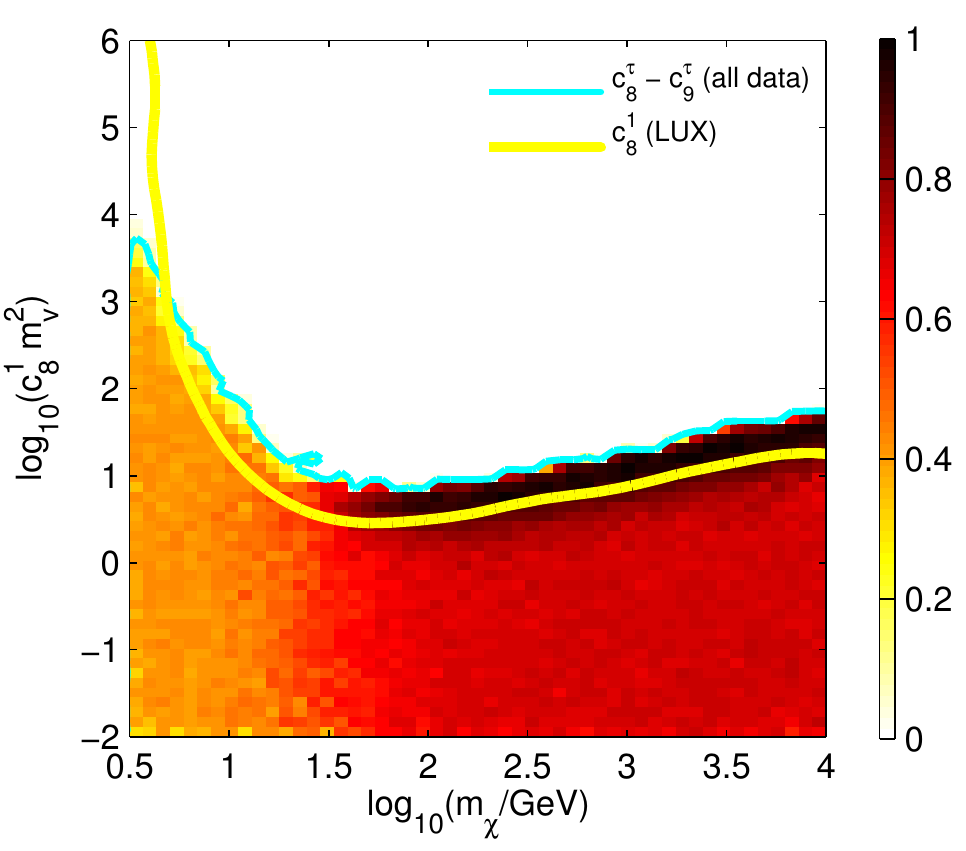}
\end{minipage}
\begin{minipage}[t]{0.49\linewidth}
\centering
\includegraphics[width=\textwidth]{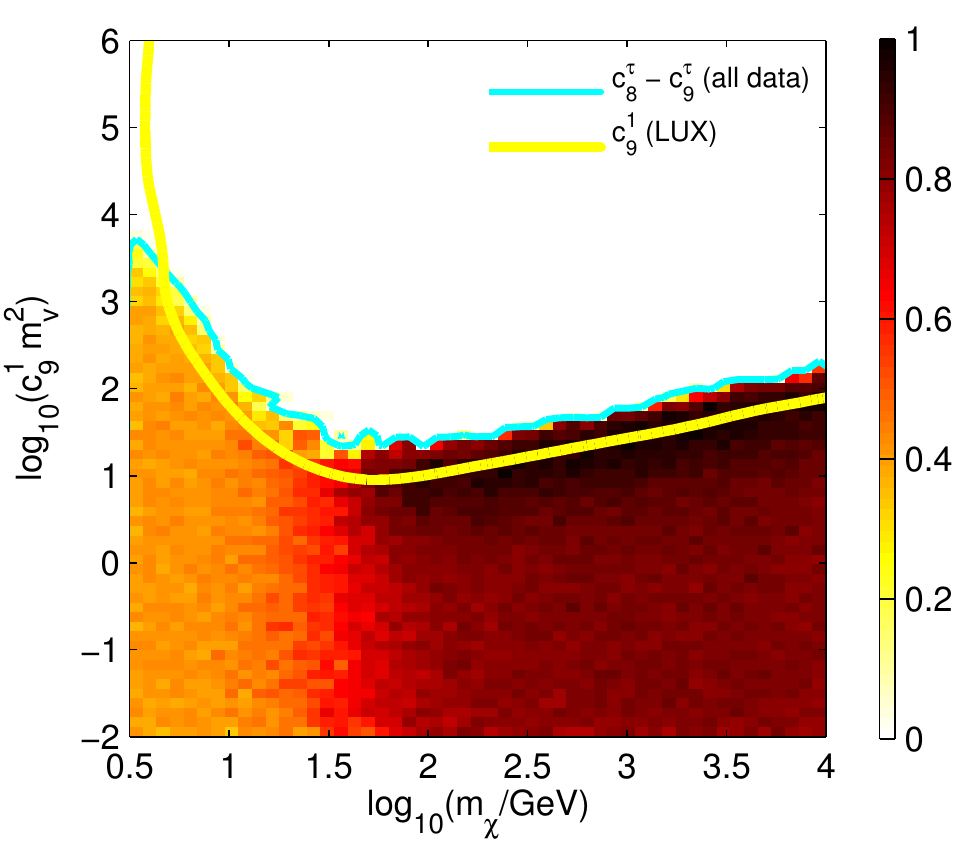}
\end{minipage}
\end{center}
\caption{Same as for Fig.~\ref{fig:c4c5c6} but now for the model parameters $c^{0}_{8}$, $c^{0}_{9}$, $c^{1}_{8}$, $c^{1}_{9}$ and $m_\chi$.}
\label{fig:c8c9}
\end{figure}

\begin{figure}[t]
\begin{center}
\begin{minipage}[t]{0.49\linewidth}
\centering
\includegraphics[width=\textwidth]{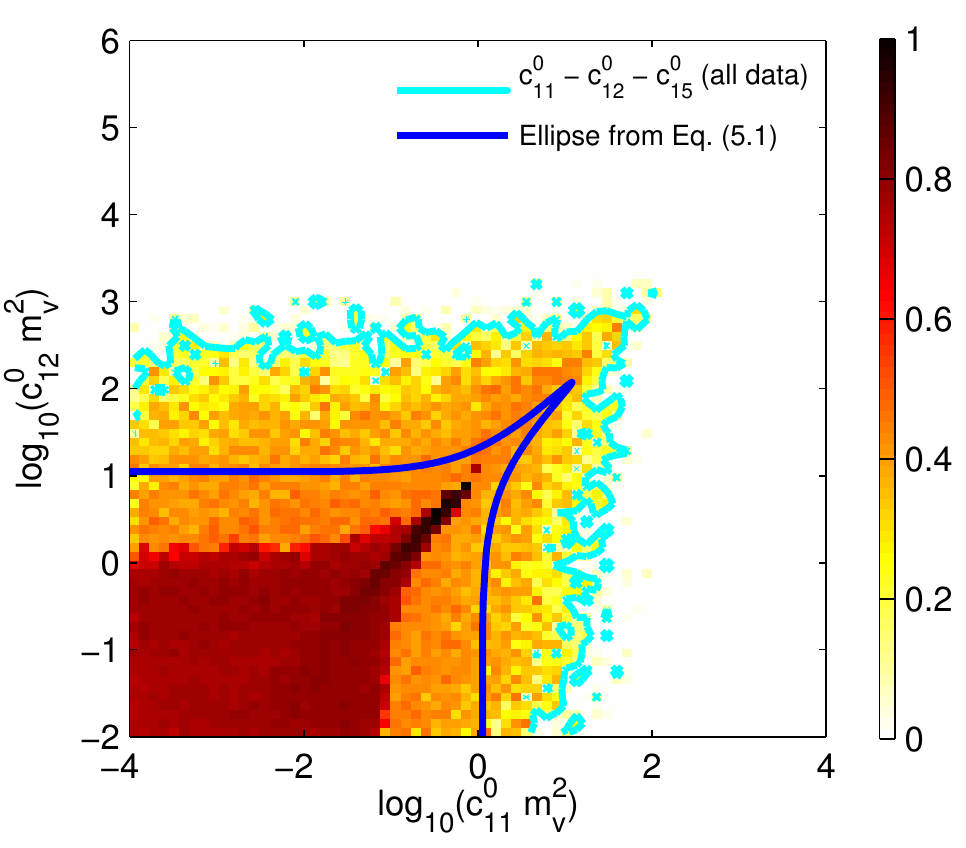}
\end{minipage}
\begin{minipage}[t]{0.49\linewidth}
\centering
\includegraphics[width=\textwidth]{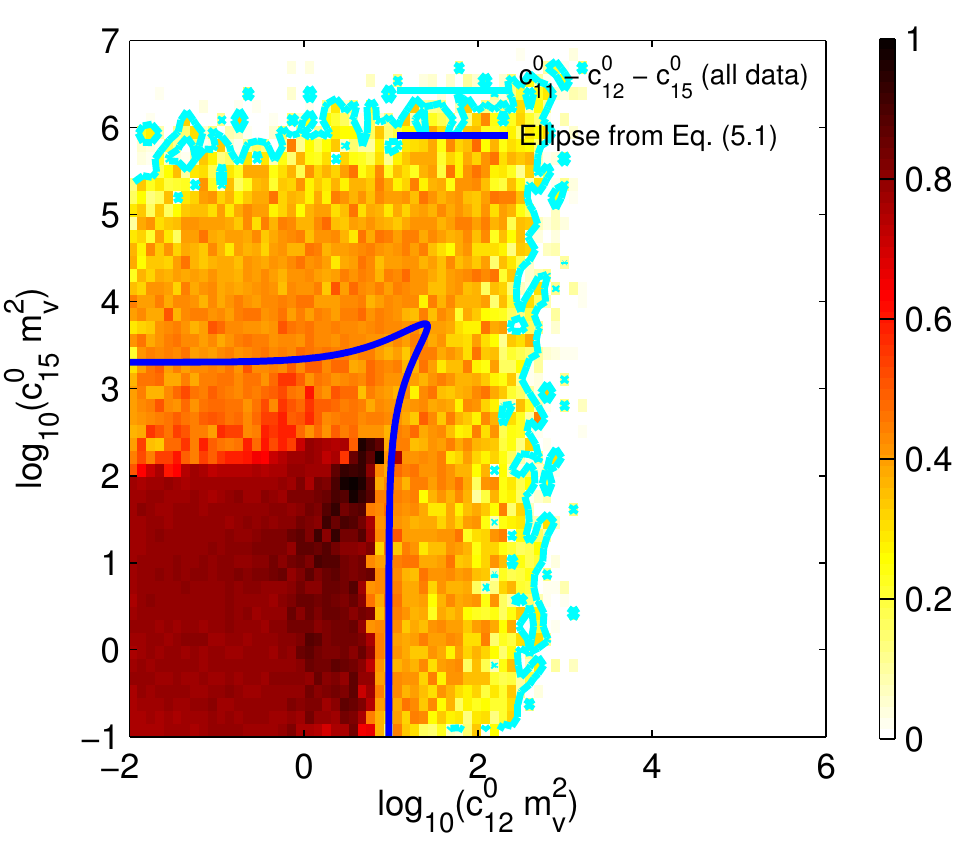}
\end{minipage}
\begin{minipage}[t]{0.49\linewidth}
\centering
\includegraphics[width=\textwidth]{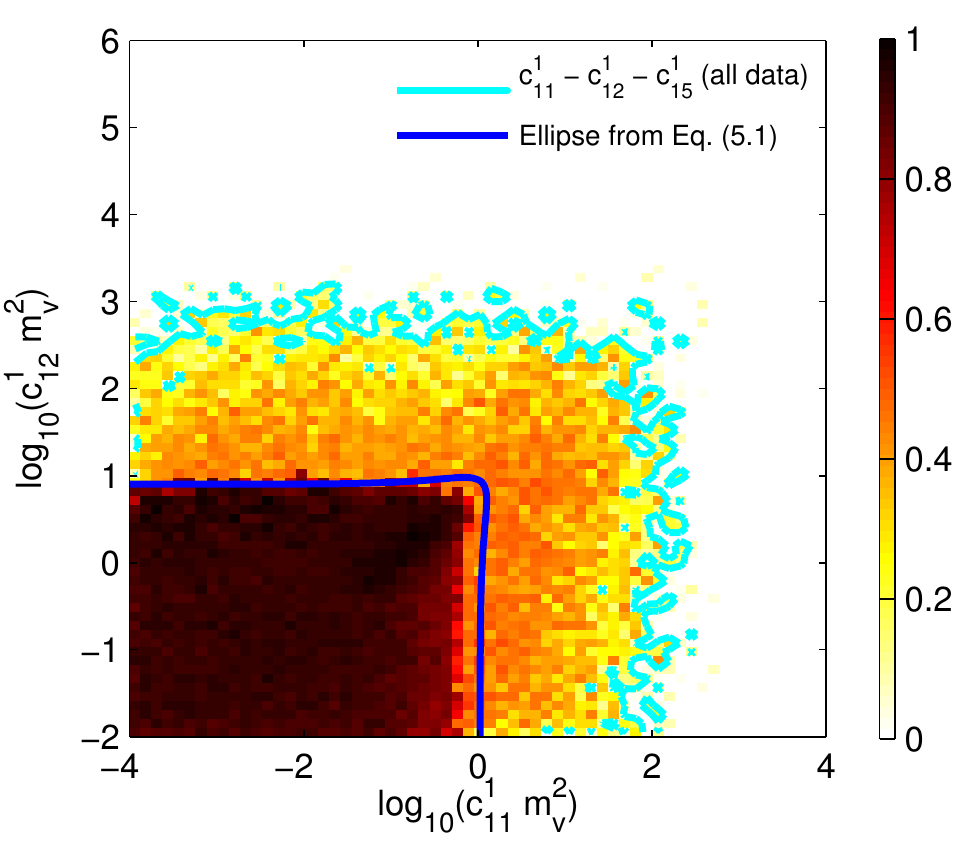}
\end{minipage}
\begin{minipage}[t]{0.49\linewidth}
\centering
\includegraphics[width=\textwidth]{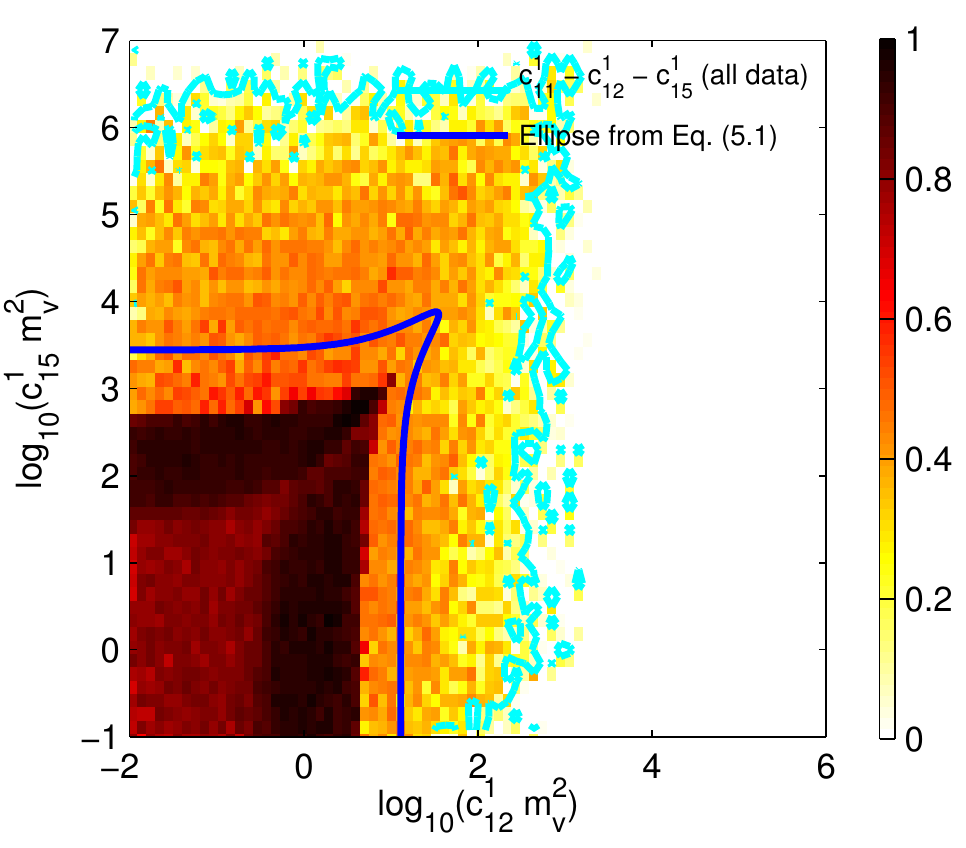}
\end{minipage}
\end{center}
\caption{Selected 2D profile Likelihoods from a fit of $m_\chi$, $\boldsymbol{\eta}$ and of the parameters in the legends to current observations. We focus on pairs of coupling constants with large correlation coefficients (see Tab.~\ref{tab:corr}). Results are presented in the planes $c_{11}^0-c_{12}^0$,  $c_{12}^0-c_{15}^0$, $c_{11}^1-c_{12}^1$,  and $c_{12}^1-c_{15}^1$. In each panel, blue lines represent ellipses drawn using Eq.~(\ref{eq:ellipse}) with the right hand side fixed at a given reference value for $m_\chi=10$~TeV, and assuming LUX as experimental setup.}
\label{fig:corr1}
\end{figure}

\begin{figure}[t]
\begin{center}
\begin{minipage}[t]{0.49\linewidth}
\centering
\includegraphics[width=\textwidth]{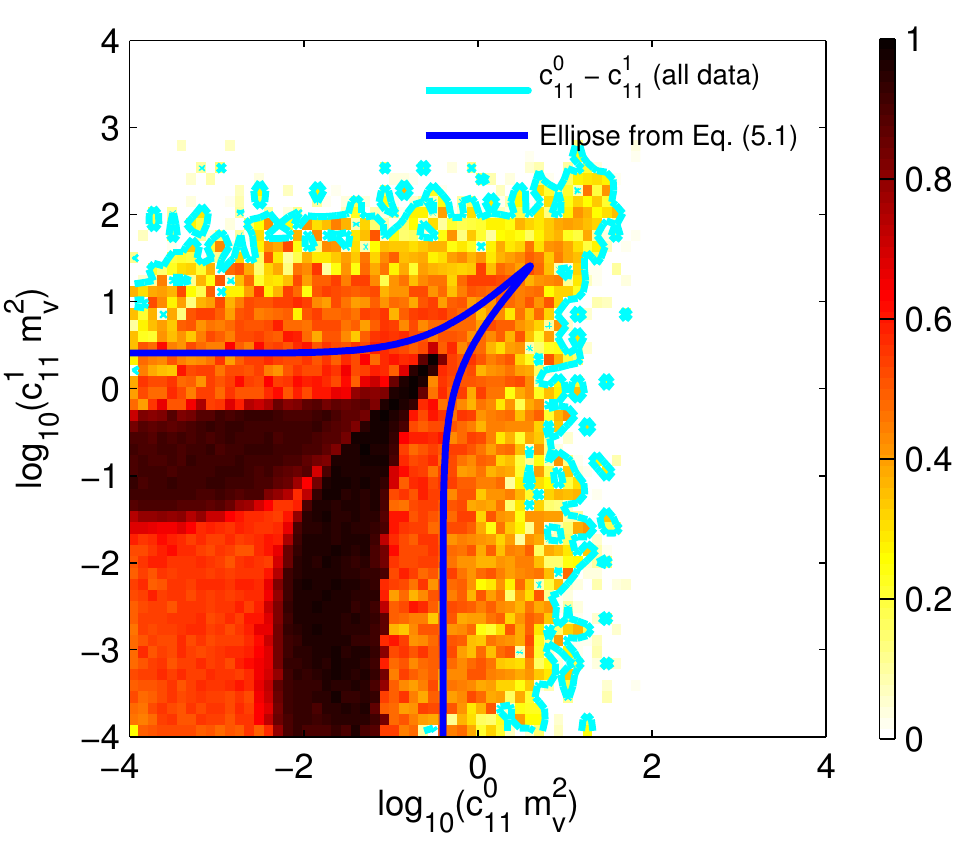}
\end{minipage}
\begin{minipage}[t]{0.49\linewidth}
\centering
\includegraphics[width=\textwidth]{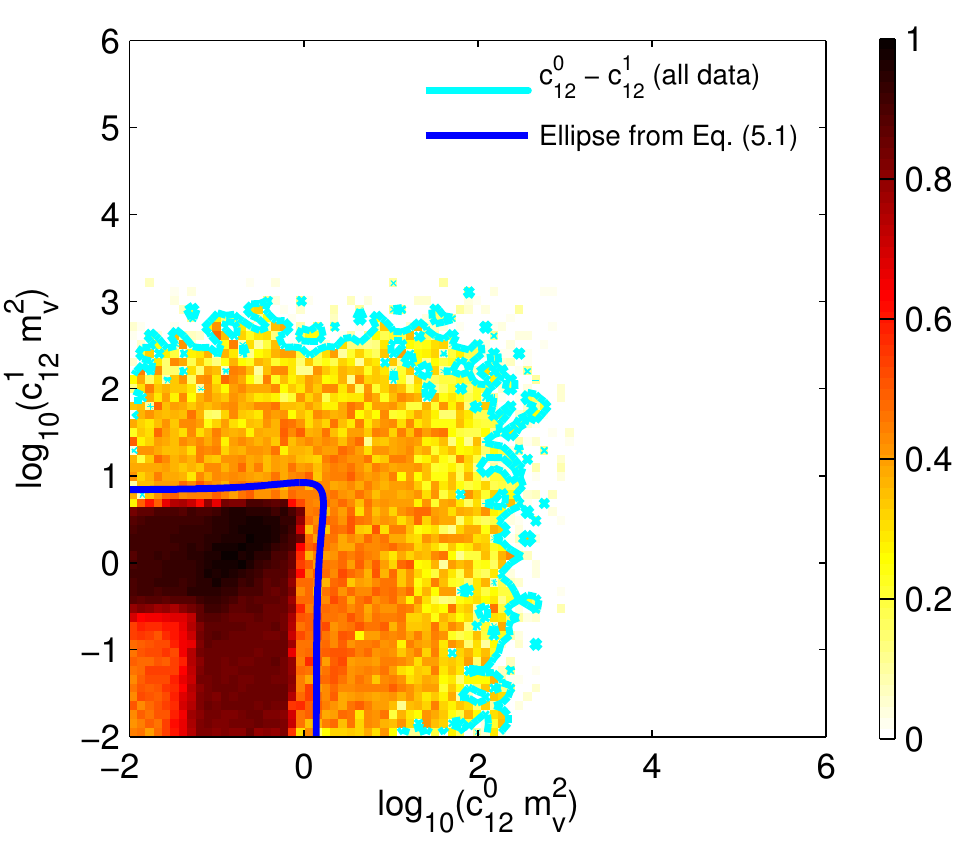}
\end{minipage}
\end{center}
\caption{Same as for Fig.~\ref{fig:corr1} but now for the model parameters $c^{0}_{11}$, $c^{1}_{11}$, $c^{0}_{12}$, and $c^{1}_{12}$.}
\label{fig:corr2}
\end{figure}

\subsection{Non-interfering operators}

We start with an analysis of the non-interfering operators. Non-interfering interaction operators are $\hat{\mathcal{O}}_7 = \hat{{\bf{S}}}_{N}\cdot {\bf{\hat{v}}}^{\perp}$, $\hat{\mathcal{O}}_{10} = i\hat{{\bf{S}}}_N\cdot{\bf{\hat{q}}}/m_N$, $\hat{\mathcal{O}}_{13} =i(\hat{{\bf{S}}}_{\chi}\cdot {\bf{\hat{v}}}^{\perp})(\hat{{\bf{S}}}_{N}\cdot{\bf{\hat{q}}}/m_N)$ and $\hat{\mathcal{O}}_{14} = i(\hat{{\bf{S}}}_{\chi}\cdot {\bf{\hat{q}}}/m_N) (\hat{{\bf{S}}}_{N}\cdot {\bf{\hat{v}}}^{\perp})$. In contrast to $\hat{\mathcal{O}}_{7}$ and $\hat{\mathcal{O}}_{10}$, the interaction operators $\hat{\mathcal{O}}_{13}$ and $\hat{\mathcal{O}}_{14}$ depend on both the momentum transfer operator and the transverse relative velocity operator. 

All non-interfering operators contribute to the nuclear spin current through the nuclear response operators $\Sigma'_{LM;\tau}$ and $\Sigma''_{LM;\tau}$, which also appear in the theory of electroweak scattering from nuclei, and characterize the familiar spin-dependent dark matter-nucleon interaction operator $\hat{\mathcal{O}}_{4}$. At the same time, the operator $\hat{\mathcal{O}}_{13}$ induces a nuclear spin-velocity current through the nuclear response operator $\tilde{\Phi}^{\prime}_{LM;\tau}$, which is specific to dark matter-nucleon interactions.

The top panels in Fig.~\ref{fig:c7c10} show the results that we obtain fitting $c_7^0$, $c_7^1$, $\boldsymbol{\eta}$ and $m_\chi$ to current direct detection experiments. Results are presented in terms of 2D profile Likelihoods (colored regions) and associated 2D 90\% confidence intervals (cyan contours) in the planes $m_\chi-c_7^0$ (left panel) and $m_\chi-c_7^1$ (right panel)\footnote{More precisely, we use the decimal logarithm of the dark matter particle mass and of the coupling constants expressed in units of $m_v^2$, with $m_v=246.2$~GeV.}. For reference, in the same panels we also report the 2D credible regions that we obtain fitting $m_{\chi}$, and a single coupling constant ($c_7^0$ in the left panel, and $c_7^1$ in the right panel)  to the LUX data. For single coupling constant fits, where marginalization is not involved, we adopt a Bayesian approach, as it requires a relatively small number of Likelihood evaluation. Volume and prior effects in dark matter direct detection are discussed in~\cite{Catena:2014uqa,Conrad:2014nna,Strege:2012kv,Arina:2011si}. The bottom panels in Fig.~\ref{fig:c7c10}, and the four panels in Fig.~\ref{fig:c13c14} show the results of similar analyses which focus on the operators $\hat{\mathcal{O}}_{10}$, $\hat{\mathcal{O}}_{13}$ and $\hat{\mathcal{O}}_{14}$.

Comparing different panels in Figs.~\ref{fig:c7c10} and~\ref{fig:c13c14}, we find that confidence intervals and profile Likelihoods in the $m_\chi-c_i^{0}$ and $m_\chi-c_i^{1}$ planes are similar for a given operator $\hat{\mathcal{O}}_i$. This result is expected, in that the (0,0) and (1,1) components of the nuclear response functions $W^{\tau\tau'}_{\Sigma'}$, $W^{\tau\tau'}_{\Sigma''}$ and $W^{\tau\tau'}_{\tilde{\Phi}'}$ are comparable for a given isotope.
The relative strength of the exclusion limits in Figs.~\ref{fig:c7c10} and~\ref{fig:c13c14} reflects the number of ${\bf{\hat{q}}}$ and ${\bf{\hat{v}}}_T^{\perp}$ operators multiplying $c_7^{\tau}$, $c_{10}^{\tau}$, $c_{13}^{\tau}$ and $c_{14}^{\tau}$ in the expressions for ${\bf{\hat{l}}}_5^{\tau}$ and ${\bf{\hat{l}}}_E^{\tau}$ in Eq.~(\ref{eq:ls}). In addition, it also depends on the relative amplitude of the nuclear response functions $W^{\tau\tau'}_{\Sigma'}$, $W^{\tau\tau'}_{\Sigma''}$ and $W^{\tau\tau'}_{\tilde{\Phi}'}$ integrated over the relevant signal regions. 

\subsection{Isoscalar-isovector interference patterns}
\label{sec:isos-isov}
For any interaction operator in Tab.~\ref{tab:operators}, including operators that do not interfere in pairs as $\hat{\mathcal{O}}_7$, $\hat{\mathcal{O}}_{10}$, $\hat{\mathcal{O}}_{13}$, and $\hat{\mathcal{O}}_{14}$, we observe isoscalar-isovector interference patterns in the rate~(\ref{rate_theory}). As an example, let us focus on the operator $\hat{\mathcal{O}}_7$. The operator $\hat{\mathcal{O}}_7$ contributes to the square modulus of the scattering amplitude as follows
\begin{equation}
\langle |\mathcal{M}_{NR}|^2\rangle_{\rm spins}^{c_7} = \frac{4\pi}{2J+1} \frac{1}{8} v_T^{\perp 2} \Bigg[ (c_7^{0})^2 \,W_{\Sigma'}^{00}(y) 
+(c_7^{1})^2 \,W_{\Sigma'}^{11}(y) +2 c_7^0 c_7^1 \,W_{\Sigma'}^{01}(y) \Bigg] \,.
\label{eq:c7int}
\end{equation}
The last term in (\ref{eq:c7int}) describes a destructive interference between the isoscalar and isovector components of the operator $\hat{\mathcal{O}}_7$, as once integrated over a typical signal region, it gives a negative contribution to the total scattering rate. We observe similar interference patterns for all operators in Tab.~\ref{tab:operators}.

In general destructive interference effects make direct detection exclusion limits weaker. 
For instance, an otherwise excluded value of $c_{7}^{0}$ can remain compatible with observations if compensated by an appropriately large value of $c_{7}^{1}$, because of the negative $\propto c_7^0 c_7^1$ term in Eq.~(\ref{eq:c7int}).

For this reason, in all figures of this work single coupling constant fits to LUX data results in stronger exclusion limits for $m_\chi\gtrsim 5$~GeV, as in these fits destructive interference effects are neglected. For smaller values of $m_\chi$, SuperCDMS, and CDMSlite dominate the exclusion limit calculation. 
 
\subsection{Interfering operators}
Interfering operators divide into 4 independent subsets. The first subset consists of the operators $\hat{\mathcal{O}}_1=\mathbb{1}_{\chi N}$ and $\hat{\mathcal{O}}_3= i\hat{{\bf{S}}}_N\cdot[({\bf{\hat{q}}}/m_N)\times{\bf{\hat{v}}}^{\perp}]$. The operator $\hat{\mathcal{O}}_1$ contributes to the nuclear vector charge through the nuclear response operator $M_{LM;\tau}$, whereas the operator $\hat{\mathcal{O}}_3$ contributes to the nuclear spin current, and to the nuclear spin-velocity current through $\Sigma'_{LM;\tau}$ and $\Phi^{\prime \prime}_{LM;\tau}$, respectively. 
The operators $\hat{\mathcal{O}}_1$ and $\hat{\mathcal{O}}_3$ generate a transition probability proportional to 
\begin{eqnarray}
\langle |\mathcal{M}_{NR}|^2\rangle_{\rm spins}^{c_1c_3} = \frac{4\pi}{2J+1}\sum_{\tau\tau'} &\Bigg[& 
c_1^{\tau}c_1^{\tau'}W_{M}^{\tau\tau'}(y) + \frac{1}{8} \frac{q^2}{m_N^2} v_T^{\perp 2}c_3^{\tau}c_3^{\tau'} W_{\Sigma'}^{\tau\tau'}(y)  \nonumber\\
&+& \frac{q^2}{m_N^2}\left( \frac{q^2}{4 m_N^2}c_3^{\tau}c_3^{\tau'} W_{\Phi''}^{\tau\tau'}(y) + c_1^{\tau}c_3^{\tau'} W_{\Phi'' M}^{\tau\tau'}(y) \right) \Bigg] \,.
\label{eq:c1c3int}
\end{eqnarray}
In Eq.~(\ref{eq:c1c3int}) terms with $\tau\neq\tau'$ describe isoscalar-isovector interference effects similar to those discussed in Sec.~\ref{sec:isos-isov}. The term $\propto c_1^{\tau}c_3^{\tau'}$ arises from the interference of $\hat{\mathcal{O}}_1$ and $\hat{\mathcal{O}}_3$.
Integrating Eq.~(\ref{eq:c1c3int}) over a typical signal region, cancellations between different terms occur, as the integrated nuclear response functions $W_{M}^{\tau\tau'}$ and $W_{\Sigma'}^{\tau\tau'}$, for $\tau\neq\tau'$, and $W_{\Phi'' M}^{\tau\tau'}$, for $\tau=\tau'$, are negative for many isotopes.

Because of cancellations in~(\ref{eq:c1c3int}), numerical noise affects the exclusion limits derived simultaneously varying $\boldsymbol{\eta}$, $m_\chi$, $c_1^0$, $c_1^1$, $c_3^0$, and $c_3^1$. To circumvent this problem, while exploring all interference patterns, here we present exclusion limits obtained in four complementary ways. In a first analysis we fit $m_\chi$, $c_1^0$, $c_3^0$, and $\boldsymbol{\eta}$ to current dark matter direct detection experiments. 
In a second analysis, we simultaneously place limits on the constants $c_1^1$ and $c_3^1$ while fitting the same data. Our third and fourth analysis  respectively consider the constants $c_1^0-c_1^1$, and $c_3^0-c_3^1$ as free parameters in the global fit. 

Fig.~\ref{fig:c1c3} shows the 2D 90\% confidence intervals (colored contours) and profile Likelihoods (colored regions) resulting from the four analyses described above. To derive the contours in the panels we fit $m_\chi$, $\boldsymbol{\eta}$, and the constants in the legends to observations. Results are presented in the $m_\chi-c_1^0$,  $m_\chi-c_1^1$,  $m_\chi-c_3^0$ and  $m_\chi-c_3^1$ planes. For comparison, in each panel we also report the 2D 90\% credible region that we obtain fitting $m_\chi$ and a single coupling constant to the LUX data.

In the top panels of Fig~\ref{fig:c1c3} (corresponding to the $m_\chi-c_1^0$ and $m_\chi-c_1^1$ planes), we observe significant differences between distinct contours. As expected, limits obtained from single coupling constant fits to LUX data are generically stronger in the $m_\chi\gtrsim 5$~GeV mass region. Global fits of the $c_1^0$ and $c_1^1$ coupling constants to current observations (cyan contours) instead produce the weakest exclusion limits, as in  Eq.~(\ref{eq:c1c3int}) cancellations between the term proportional to $c_1^\tau c_1^\tau$, with $\tau=0$ or  $\tau=1$, and the one proportional to $c_1^0 c_1^1$ can be large. In contrast, distinct contours in the bottom panels of Fig~\ref{fig:c1c3} do not significantly differ, as in Eq.~(\ref{eq:c1c3int}) cancellations between the term $\propto c_3^\tau c_3^\tau$, with $\tau=0$ or  $\tau=1$, and the terms $\propto c_1^{\tau} c_3^{\tau}$ or $\propto c_3^{\tau}c_3^{\tau'}$, with $\tau\neq\tau'$,  are less important.  

The second subset of interfering operators consists of $\hat{\mathcal{O}}_{11}=i{\bf{\hat{S}}}_\chi\cdot ({\bf{\hat{q}}}/m_N)$, $\hat{\mathcal{O}}_{12}= \hat{{\bf{S}}}_{\chi}\cdot (\hat{{\bf{S}}}_{N} \times{\bf{\hat{v}}}^{\perp})$, and $\hat{\mathcal{O}}_{15}= -[\hat{{\bf{S}}}_{\chi}\cdot ({\bf{\hat{q}}}/m_N)][ (\hat{{\bf{S}}}_{N}\times {\bf{\hat{v}}}^{\perp} ) \cdot ({\bf{\hat{q}}}/m_N)]$. The operator  $\hat{\mathcal{O}}_{11}$ contributes to the nuclear vector charge through the nuclear response operator $M_{LM;\tau}$, similarly to $\hat{\mathcal{O}}_{1}$. The operators $\hat{\mathcal{O}}_{12}$ and $\hat{\mathcal{O}}_{15}$ induce a nuclear spin current through $\Sigma'_{LM;\tau}$ and $\Sigma''_{LM;\tau}$, and a nuclear spin-velocity current through $\Phi''_{LM;\tau}$.

The coupling constants $c_{11}^{\tau}$, $c_{12}^{\tau}$, and $c_{15}^{\tau}$ contribute to the scattering amplitude in a way similar to $c_1^\tau$ and $c_3^\tau$ in Eq.~(\ref{eq:c1c3int}). Also in this case, we therefore expect cancellations in the scattering rate due to destructive interference between operator pairs, or between isoscalar and isovector components of the same operator. To study the impact of multi-interaction interference effects in the exclusion limit calculation, we proceed as for the operators $\hat{\mathcal{O}}_1$ and $\hat{\mathcal{O}}_3$. We therefore divide the coupling constants $c_{11}^{\tau}$, $c_{12}^{\tau}$, and $c_{15}^{\tau}$ into 5 subsets: $(c_{11}^{0}, c_{12}^{0}, c_{15}^{0})$, $(c_{11}^{1}, c_{12}^{1}, c_{15}^{1})$ and $(c_{i}^{0}, c_{i}^{1})$, with $i=11,12,15$. We then fit $m_\chi$, $\boldsymbol{\eta}$, and a single subset of coupling constants at the time to observations. Finally, from each fit we derive exclusion limits on the coupling constants in analysis. This procedure allows to separate interference effects due to operator pairs, from those related to isoscalar and isovector components of the same operator. 

Fig.~\ref{fig:c11c12c15} shows the 2D 90\% confidence intervals and profile Likelihoods that we find in the five analyses described above. Results are presented in 6 $m_\chi$ vs coupling constant planes. To derive the contours in the panels we fit $m_\chi$, $\boldsymbol{\eta}$, and the constants in the legends to observations. For comparison, in each panel we also report the 2D 90\% credible regions of a single coupling constant fit to the LUX data. In the $m_\chi - c_{12}^1$, $m_\chi - c_{15}^0$ and $m_\chi - c_{15}^1$ planes  distinct exclusion limits are comparable. In the other panels, contours differ by up to 1 order of magnitude in the coupling constants. As for $\hat{\mathcal{O}}_1$ and $\hat{\mathcal{O}}_3$, such differences arise from multi-interaction interference effects.

There are two additional subsets of interfering operators. One consists of the operators $\hat{\mathcal{O}}_4 =  \hat{{\bf{S}}}_{\chi}\cdot \hat{{\bf{S}}}_{N}$,  $\hat{\mathcal{O}}_5 = i{\bf{\hat{S}}}_\chi\cdot[({\bf{\hat{q}}}/m_N)\times{\bf{\hat{v}}}^{\perp}]$, and $\hat{\mathcal{O}}_6 = [{\bf{\hat{S}}}_\chi\cdot({\bf{\hat{q}}}/m_N)] [\hat{{\bf{S}}}_N\cdot(\hat{{\bf{q}}}/m_N)]$. The other one includes $\hat{\mathcal{O}}_8 = \hat{{\bf{S}}}_{\chi}\cdot {\bf{\hat{v}}}^{\perp}$ and $\hat{\mathcal{O}}_9 = i{\bf{\hat{S}}}_\chi\cdot[\hat{{\bf{S}}}_N\times({\bf{\hat{q}}}/m_N)]$. Eq.~(\ref{eq:ls}) shows the contribution of these operators to the nuclear charges and currents. The operators $\hat{\mathcal{O}}_5$ and $\hat{\mathcal{O}}_8$ are the only operators contributing to the nuclear convection current (through $\Delta_{LM;\tau}$). 
Figs.~\ref{fig:c4c5c6} and \ref{fig:c8c9} show the 2D 90\% confidence intervals and profile Likelihoods that we find analyzing the operators $\hat{\mathcal{O}}_4$, $\hat{\mathcal{O}}_5$, $\hat{\mathcal{O}}_6$, $\hat{\mathcal{O}}_8$, and $\hat{\mathcal{O}}_9$. In these figures we do not observe strong destructive interference patterns.

We conclude this section with an analysis of selected pairs of coupling constants. We focus on largely correlated coupling constants (see Tab.~\ref{tab:corr}). Figs.~\ref{fig:corr1} and \ref{fig:corr2} show the 2D profile Likelihoods that we obtain fitting $m_\chi$, $\boldsymbol{\eta}$ and  the parameters in the legends to current observations. Results are presented in the planes $c_{11}^0-c_{12}^0$,  $c_{12}^0-c_{15}^0$, $c_{11}^1-c_{12}^1$,  $c_{12}^1-c_{15}^1$, $c_{11}^0-c_{11}^1$, and $c_{12}^0-c_{12}^1$. In each panel, blue lines represent ellipses constructed from Eq.~(\ref{eq:ellipse}) with the right hand side fixed at a given reference value, and assuming LUX as experimental setup. As expected, the red elliptical regions in Figs.~\ref{fig:corr1} and \ref{fig:corr2} are in general aligned with the blue ellipses. However, the 2D profile Likelihood in the $c_{12}^0-c_{15}^0$ plane constitutes an exception. The mismatch observed in this figure is due to the fact that the blue ellipse in the top right panel of Fig.~\ref{fig:corr1} neglects a large $c_{11}^0-c_{12}^0$ correlation, which is instead taken into account deriving the corresponding red elliptical region.

\section{Conclusions}
\label{sec:conclusions}
We compared the general effective theory of one-body dark matter-nucleon interactions to current direct detection experiments in a global multidimensional statistical analysis. In this study, we included data from a variegated sample of direct detection experiments in a single Likelihood function. In the fits we simultaneously varied subsets of correlated model parameters, covering a large volume of the parameter space. 
We presented our results in terms of 90\% confidence level exclusion limits on the 28 isoscalar and isovector coupling constants of the theory. 

We found that current direct detection experiments are able to probe all isoscalar and isovector coupling constants, including those measuring the strength of dark matter-nucleon interactions commonly neglected in this context. For instance, the interaction operators $\hat{\mathcal{O}}_{11} = i{\bf{\hat{S}}}_\chi\cdot ({\bf{\hat{q}}}/m_N)$ and $\hat{\mathcal{O}}_{12} = \hat{{\bf{S}}}_{\chi}\cdot (\hat{{\bf{S}}}_{N} \times{\bf{\hat{v}}}^{\perp} )$, and the familiar spin-dependent interaction operator $\hat{\mathcal{O}}_4 = \hat{{\bf{S}}}_{\chi}\cdot \hat{{\bf{S}}}_{N}$ are equally constrained by current data, though the former are by far less explored.

We characterized the interference patterns arising in dark matter direct detection from multi-interaction effects involving pairs of dark matter-nucleon interaction operators, or isoscalar and isovector components of the same operator. 
Most importantly, we quantified the impact of multi-interaction interference effects on the calculation of direct detection exclusion limits.

We found that destructive interference effects weaken direct detection exclusion limits derived neglecting the superposition of different interaction operators by up to one order of magnitude. Destructive interference effects largely affect exclusion limits on the strength of the familiar spin-independent interaction $\hat{\mathcal{O}}_{1}$, and are less important for the standard spin-dependent operator $\hat{\mathcal{O}}_{4}$.  

\acknowledgments R.~C. acknowledges partial support from a start-up grant funded by the University of G\"ottingen, and from the European Union FP7 ITN INVISIBLES (Marie Curie Actions, PITN-GA-2011-289442). P.~G. has partially been supported by NSF award PHY-1415974.

\appendix
\section{Dark matter response functions}
\label{sec:appDM}
Below, we list the dark matter response functions that appear in Eq.~(\ref{rate_theory}). The notation is the same used in the body of the paper. In the figures, for definitiveness we assume $j_\chi=1/2$.
\begin{eqnarray}
 R_{M}^{\tau \tau^\prime}\left(v_T^{\perp 2}, {q^2 \over m_N^2}\right) &=& c_1^\tau c_1^{\tau^\prime } + {j_\chi (j_\chi+1) \over 3} \left[ {q^2 \over m_N^2} v_T^{\perp 2} c_5^\tau c_5^{\tau^\prime }+v_T^{\perp 2}c_8^\tau c_8^{\tau^\prime }
+ {q^2 \over m_N^2} c_{11}^\tau c_{11}^{\tau^\prime } \right] \nonumber \\
 R_{\Phi^{\prime \prime}}^{\tau \tau^\prime}\left(v_T^{\perp 2}, {q^2 \over m_N^2}\right) &=& {q^2 \over 4 m_N^2} c_3^\tau c_3^{\tau^\prime } + {j_\chi (j_\chi+1) \over 12} \left( c_{12}^\tau-{q^2 \over m_N^2} c_{15}^\tau\right) \left( c_{12}^{\tau^\prime }-{q^2 \over m_N^2}c_{15}^{\tau^\prime} \right)  \nonumber \\
 R_{\Phi^{\prime \prime} M}^{\tau \tau^\prime}\left(v_T^{\perp 2}, {q^2 \over m_N^2}\right) &=&  c_3^\tau c_1^{\tau^\prime } + {j_\chi (j_\chi+1) \over 3} \left( c_{12}^\tau -{q^2 \over m_N^2} c_{15}^\tau \right) c_{11}^{\tau^\prime } \nonumber \\
  R_{\tilde{\Phi}^\prime}^{\tau \tau^\prime}\left(v_T^{\perp 2}, {q^2 \over m_N^2}\right) &=&{j_\chi (j_\chi+1) \over 12} \left[ c_{12}^\tau c_{12}^{\tau^\prime }+{q^2 \over m_N^2}  c_{13}^\tau c_{13}^{\tau^\prime}  \right] \nonumber \\
   R_{\Sigma^{\prime \prime}}^{\tau \tau^\prime}\left(v_T^{\perp 2}, {q^2 \over m_N^2}\right)  &=&{q^2 \over 4 m_N^2} c_{10}^\tau  c_{10}^{\tau^\prime } +
  {j_\chi (j_\chi+1) \over 12} \left[ c_4^\tau c_4^{\tau^\prime} + \right.  \nonumber \\
 && \left. {q^2 \over m_N^2} ( c_4^\tau c_6^{\tau^\prime }+c_6^\tau c_4^{\tau^\prime })+
 {q^4 \over m_N^4} c_{6}^\tau c_{6}^{\tau^\prime } +v_T^{\perp 2} c_{12}^\tau c_{12}^{\tau^\prime }+{q^2 \over m_N^2} v_T^{\perp 2} c_{13}^\tau c_{13}^{\tau^\prime } \right] \nonumber \\
    R_{\Sigma^\prime}^{\tau \tau^\prime}\left(v_T^{\perp 2}, {q^2 \over m_N^2}\right)  &=&{1 \over 8} \left[ {q^2 \over  m_N^2}  v_T^{\perp 2} c_{3}^\tau  c_{3}^{\tau^\prime } + v_T^{\perp 2}  c_{7}^\tau  c_{7}^{\tau^\prime }  \right]
       + {j_\chi (j_\chi+1) \over 12} \left[ c_4^\tau c_4^{\tau^\prime} +  \right.\nonumber \\
       &&\left. {q^2 \over m_N^2} c_9^\tau c_9^{\tau^\prime }+{v_T^{\perp 2} \over 2} \left(c_{12}^\tau-{q^2 \over m_N^2}c_{15}^\tau \right) \left( c_{12}^{\tau^\prime }-{q^2 \over m_N^2}c_{15}^{\tau \prime} \right) +{q^2 \over 2 m_N^2} v_T^{\perp 2}  c_{14}^\tau c_{14}^{\tau^\prime } \right] \nonumber \\
     R_{\Delta}^{\tau \tau^\prime}\left(v_T^{\perp 2}, {q^2 \over m_N^2}\right)&=&  {j_\chi (j_\chi+1) \over 3} \left[ {q^2 \over m_N^2} c_{5}^\tau c_{5}^{\tau^\prime }+ c_{8}^\tau c_{8}^{\tau^\prime } \right] \nonumber \\
 R_{\Delta \Sigma^\prime}^{\tau \tau^\prime}\left(v_T^{\perp 2}, {q^2 \over m_N^2}\right)&=& {j_\chi (j_\chi+1) \over 3} \left[c_{5}^\tau c_{4}^{\tau^\prime }-c_8^\tau c_9^{\tau^\prime} \right].
\label{eq:R}
\end{eqnarray}

\providecommand{\href}[2]{#2}\begingroup\raggedright\endgroup

\end{document}